\documentclass[12pt]{article}
\usepackage[dvips]{color}                 
\usepackage{graphicx}                     
\usepackage{amssymb}
\usepackage{xspace}                       
\renewcommand{\today}{\ifcase\day\or 1st\or 2nd\or 3rd\or 4th\or 5th\or 6th\or

        7th\or 8th\or 9th\or 10th\or 11th\or 12th\or 13th\or 14th\or 15th\or 

        16th\or 17th\or 18th\or 19th\or 20th\or 21st\or 22nd\or 23rd\or 24th\or

        25th\or 26th\or 27th\or 28th\or 29th\or 30th\or 

        31st\fi~\ifcase\month\or January\or February\or March\or April\or

        May\or June\or July\or August\or September\or October\or November\or

        December\fi \space \number\year}   
\newcommand{\Tr}{\mathrm{Tr}}

\newcommand{\fpi}{f_\pi}

\newcommand{\mytitle}[1]{
                         \begin{center}
                           \LARGE{\textbf{#1}}
                         \end{center}}
\newcommand{\myauthor}[1]{\textbf{#1}}
\newcommand{\myaddress}[1]{\textit{#1}}
\newcommand{\mypreprint}[1]{\begin{flushright} #1 \end{flushright}}

\begin{document}

%

\begin{titlepage}
  \mypreprint{TUM-T39-02-05  
\hfill
    ECT* 02-08 \\
    March 31, 2002}


  \mytitle{Chiral Magnetism of the Nucleon\footnote{Work supported in part by BMBF and DFG}}


\begin{center}
  \myauthor{Thomas R. Hemmert$^{a,c}$}
  and
  \myauthor{Wolfram Weise$^{a,b}$} 

  \vspace*{0.5cm}

  \myaddress{$^a$
    Physik-Department, Theoretische Physik \\
    Technische Universit{\"a}t M{\"u}nchen, D-85747 Garching, Germany\\
    (Email: themmert@physik.tu-muenchen.de)}\\[2ex]
  \myaddress{$^b$ ECT*, Villa Tambosi, I-38050 Villazzano (Trento), Italy\\
    (Email: weise@ect.it)}\\[2ex]
  \myaddress{$^c$ Nuclear Theory Group\footnote{Visiting Scholar}, 
               Department of Physics\\
               University of Washington, Seattle, WA 98195, USA}

  \vspace*{0.2cm}
\end{center}

\vspace*{0.5cm}

\begin{abstract}
  We study the quark mass expansion of the magnetic moments of the nucleon
in a chiral effective field theory including nucleons, pions and delta resonances as explicit degrees of freedom. We point out that the usual powercounting applied so far to this problem misses important quark mass structures generated via an intermediate isovector M1 nucleon-delta transition. We propose a modified powercounting and compare the resulting chiral extrapolation function to available (quenched) lattice data. The extrapolation is found to work surprisingly well, given that the lattice data result from rather large quark masses. Our calculation raises the hope that extrapolations of lattice data utilizing chiral effective field theory might be applicable over a wider range in quark masses than previously thought, and we discuss some open questions in this context. Furthermore, we observe that within the current lattice data uncertainties the extrapolations presented here are consistent with the Pade fit ansatz introduced by the Adelaide group a few years ago. 
\end{abstract}

\vskip 1.0cm

\noindent
\begin{tabular}{rl}
Suggested PACS numbers:& \\[1ex]
Suggested Keywords: &\begin{minipage}[t]{11cm}
                   Effective Field Theory, Lattice QCD, Chiral Extrapolation 
                    \end{minipage}
\end{tabular}

\vskip 1.0cm
\end{titlepage}

\setcounter{page}{2} \setcounter{footnote}{0} \newpage

%


\section{Introduction}
\setcounter{equation}{0}


The computation of nucleon properties in Lattice QCD is progressing with
steadily increasing accuracy \cite{recentlattice}. So far, these results are,
however, limited to relatively large quark masses and the quenched
approximation. The typical ``light'' quark masses manageable on the lattice up
to now, are more than 10-20 times larger than the average u- and d-quark
masses, $m_q\sim 8$ MeV, determined at a renormalization scale around 1
GeV. This corresponds to pion masses well above 0.5 GeV.

On the other hand, in the chiral limit $m_q\rightarrow 0$, QCD at low energies is realized in the form of an effective field theory with spontaneously broken chiral symmetry, with massless pions as the primary active degrees of freedom. Nucleons are added to this theory as fermionic matter fields which can be treated non-relativistically due to their large mass. The coupling of the chiral Goldstone bosons (i.e. the pions) to these spin 1/2 matter fields (i.e. the nucleons) produces the so called ``pion-cloud'' of the nucleon, an important component of nucleon structure at low energy and momentum scales (for a survey and references see \cite{TW}).

The interpolation between lattice QCD results and the chiral limit, passing through the physical pion mass point at $m_\pi\approx 0.14$ GeV, is presently a subject of intense studies and lively debate. Leinweber et al. \cite{adelaide} have initiated such considerations by requiring that any interpolation of this kind should be subject to the leading dependence on the pion mass, as dictated by chiral symmetry.

In the present work\footnote{Some aspects of this work have already been reported in Refs.\cite{Nstar,baryons}.} 
we focus on the chiral aspects of nucleon magnetic
moments. The authors of Ref.\cite{adelaide} have introduced a parameterized
form of the pion mass dependence of proton and neutron magnetic moments in
order to interpolate between available (quenched) lattice QCD results and the
``physical'' values at the proper $m_\pi$. Our approach uses instead systematic
methods of chiral perturbation theory (ChPT) in the baryon sector \cite{TW}. It will be
demonstrated that such a scheme provides an interpolating function of $m_\pi$,
which successfully connects lattice data with actual physical magnetic moments,
in such a way that the physics behind the chiral magnetic structure of the
nucleon can be identified. Not surprisingly, virtual excitations of the
delta isobar turn out to play a decisive role in this context. The strong
$N\rightarrow\Delta$ magnetic dipole transition induced by the photon field is
a prominent feature in determining the non-linear dependence of the magnetic
moments on the pion mass. Given that delta excitations are separated from
the nucleon mass by a ``small'' scale of hardly more than twice $m_\pi$, it is
obviously not a good starting point to relegate the important physics of the
$\Delta$(1232) to higher order ``counter terms'', if one is interested in the
$m_\pi$-dependence of nucleon structure properties {\it above} the physical
pion mass. Such effective chiral field theories with explicit pion, nucleon and
delta degrees of freedom and a systematic power counting already exist in the
literature (e.g. see Ref.\cite{review} and references therein). Here, however,
we argue that even these existing schemes, such as the so-called ``Small Scale Expansion'' of Ref.\cite{review}, need to be modified if one wants to capture the important $\Delta$(1232) induced $m_\pi$-dependence already in a leading one-loop calculation.

This paper is organized as follows. The chiral effective field theory
framework, including the treatment of nucleon and delta degrees of freedom,
will be summarized in section \ref{Lagrangians}. 
The actual calculation of the anomalous
magnetic moments is described in section \ref{calculation}, with the results discussed in
sections \ref{results}, \ref{numerik} and \ref{isoscalar}. The following two sections examine higher order corrections to be
compared with the Pade-approximants suggested by the Adelaide group
\cite{adelaide}, and discuss some aspects of the quenched approximation as they
may be relevant in extrapolations toward the presently existing lattice
data. Undeniably, uncertainties still remain in such extrapolations to pion
masses as large as five times the physical one. The results are nevertheless
promising, and further progress in lattice QCD towards smaller quark masses in
partially-quenched or even unquenched simulations can be expected to reduce the errors substantially.

\section{Effective Field Theory Input}\label{Lagrangians}
\setcounter{equation}{0}

\subsection{General Remarks}\label{general}

The effective field theory that approximates QCD in the low-energy limit is
Chiral Perturbation Theory (ChPT) \cite{weinberg,GL}. In the case of two light
quark flavors its primary degrees of
freedom are pions as Goldstone bosons of spontaneously broken chiral
symmetry. ChPT represents a systematic expansion in terms of low-momentum and
small quark mass scales, which allows for a systematic treatment of the
additional explicit breaking of chiral symmetry responsible for non-zero pion
mass. Baryons are introduced as fermionic matter fields, 
acting as sources which
create or annihilate pion fields in accordance with chiral symmetry. In 
the non-relativistic version\footnote{Some results of relativistic baryon ChPT are discussed in appendix \ref{B}.} of baryon ChPT considered in this work (``heavy baryon ChPT'') \cite{TW}, the nucleon mass $M$ is treated
as a ``large'' scale that persists in the chiral limit, allowing for a $1/M$
expansion in the theory. 

The $\Delta$(1232) isobar is the lowest spin-3/2
excitation of the nucleon, reached via a strong magnetic dipole transition. It
is therefore of key importance in considerations of the nucleon's magnetic
structure. Since its mass differs from that of the nucleon by less than 0.3
GeV, its incorporation as an explicit degree of freedom in baryon ChPT
turns out to be
mandatory in the present context. The basic techniques to do so are well established, at least in the context of non-relativistic baryon ChPT \cite{JM}. 
However, we emphasize that a chiral effective field theory that couples both
octet and decuplet baryons to the Goldstone boson dynamics in addition must
specify how the small but {\em always finite} octet-decuplet
mass-splitting---throughout this work denoted by the parameter $\Delta$---is
merged with the traditional powers-of-$q$ counting of ChPT spelled out in
Ref.\cite{GL}. In that sense we state that a {\em unique extension} of spin 1/2
baryon ChPT to a chiral effective field theory with explicit spin 1/2 and spin
3/2 baryon degrees of freedom does {\em not} exist. An additional piece of
information concerning the proper (ac)counting of $\Delta$ has to be provided,
without guidance from chiral symmetry constraints\footnote{For example, we note
that a combined treatment of non-relativistic baryon ChPT and large
$N_c$-counting rules can provide a systematic power counting for systems with
both spin 1/2 and spin 3/2 degrees of freedom \cite{DJM}.}. Here we
specifically address this issue by adopting the (phenomenologically motivated)
philosophy, spelled out in Ref.\cite{letter}, to count in the parameter
$\epsilon\equiv(q,\Delta)$ which, in addition to the (usual) powers-of-$q$, also keeps track of the finite mass-splitting $\Delta$. All discussions in the subsequent chapters refering to leading order (LO), next-to-leading order (NLO), etc. Lagrangians are then understood as ``powers-of-$\epsilon$'', i.e. ${\cal O}(\epsilon),\,{\cal O}(\epsilon^2),\ldots$ \cite{review}. 

The following subsections give a brief summary of our basic input. We will work
with 2-flavor ChPT coupled to spin 1/2 and spin 3/2 matter fields throughout.
We also note that strictly speaking all parameters $g_i,\,m_i,\,f_i,\dots$ in the Lagrangians to be introduced in section \ref{Lagrangians} should carry an extra superscript index $g_i^0,\,m_i^0\ldots$ to distinguish them from any physical quantities which might carry the same label. In most cases we suppress this extra index to obtain simpler formulae, but it is understood that there is a difference, even if it sometimes is of higher order than considered in this calculation.

\subsection{Pion Lagrangian}

For one-loop calculations of the nucleon magnetic moments, the SU(2) chiral 
Lagrangian for pion fields ${\bf  \pi}^a\;(a=1,2,3)$ in the
 presence of an external electromagnetic field $A_\mu$ is needed only to leading 
order. With $e$ denoting the unit charge we utilize \cite{GL}
\begin{eqnarray}
{\cal L}_{\pi\pi}&=&\frac{\fpi^2}{4}\;Tr\left[\nabla_\mu U^\dagger\nabla^\mu 
U+\chi^\dagger U+\chi U^\dagger\right] ,
\end{eqnarray}
with the chiral tensors
\begin{eqnarray}
U&=&\sqrt{1-\frac{{\bf \vec{\pi}}^{\,2}}{\fpi^2}}+\frac{i}{\fpi}{\bf \vec{\tau}
\cdot\vec{\pi}}\nonumber\\
\nabla_\mu U&=&\partial_\mu U-i\,F_\mu^R\,U+i\,U\,F_\mu^L\nonumber\\
\chi&=&2 B_0 {\cal M}+\ldots
\end{eqnarray}
Here $\fpi$ represents the pion-decay constant (in the chiral limit)  and
$\tau^a$ are the usual SU(2) (Pauli) isospin matrices. As we are only interested in
interactions of $u$ and $d$ quarks with an electromagnetic background field, 
the left- and right-handed
(axial) vector source terms can be identified as 
\begin{eqnarray}
F_\mu^L=F_\mu^R=\frac{e}{2}\,A_\mu\,\tau^3 \, .
\end{eqnarray}
Furthermore, ${\cal M}$ denotes the quark
mass matrix for the case of 2 light flavors $u,d$. The connection 
between the
chiral condensate parameter $B_0$, the non-zero quark masses and the resulting 
non-zero masses for the pion fields will be discussed later.

\subsection{Nucleon and Delta Lagrangians}
\subsubsection{Leading Order Lagrangians}

The leading order $N\pi$ and $\Delta\pi$ Lagrangians (diagonal in $N$ and
$\Delta$)
required for a calculation 
with explicit pion, nucleon and delta degrees of freedom in 
the presence of external electromagnetic fields are \cite{review,JM}:
\begin{eqnarray}
{\cal L}^{(1)}_{N}&=&\bar{N}_v\left[i\,v\cdot D+g_A\,S\cdot u\right]N_v
\; ,\nonumber\\
{\cal L}^{(1)}_{\Delta}&=&-\,\bar{T}_i^\mu\left[i\,v\cdot D^{ij}-
\xi^{ij}_{I=3/2}\Delta+g_1\,S\cdot u^{ij}\right]g_{\mu\nu}\,T_j^\nu\; , 
\label{eq:lag1}
\end{eqnarray}
where $N_v$ corresponds to the non-relativistic spin 1/2 nucleon field and 
$T_i^\mu$ denotes the non-relativistic spin 3/2 delta field with 
free 4-vector index $\mu$ (in Rarita-Schwinger notation) and an isovector 
index 
$i$ (in isospurion notation) \cite{review}. The 
4-velocity vector $v^\mu$ occurs in the non-relativistic reduction of the 
fully 
Lorentz-invariant chiral Lagrangians, and $S^\mu$ denotes the 
Pauli-Lubanski spin-vector---details on calculating with non-relativistic 
chiral effective field theories in the so-called ``heavy baryon'' regime 
can be found in \cite{BKM}. The chiral tensors needed for the  calculation of 
the anomalous magnetic moments read
\begin{eqnarray}
D_\mu&=&\partial_\mu+\Gamma_\mu-iV_\mu^{(s)}\,,\nonumber\\
\Gamma_\mu&=&\frac{1}{2}\left[\sqrt{U}^{\,\dagger},\partial_\mu\sqrt{U}\right]
-\frac{i}{2}\sqrt{U}^{\,\dagger} F_\mu^R\sqrt{U}
-\frac{i}{2}\sqrt{U}F_\mu^L\sqrt{U}^{\,\dagger}\,,\nonumber\\
u_\mu&=&i\sqrt{U}^{\,\dagger}\,\nabla_\mu U\,\sqrt{U}^{\,\dagger}\,,\quad\quad
u_\mu^{ij}\;=\;u_\mu\,\delta^{ij}\,,\nonumber\\
D_\mu^{ij}&=&\partial_\mu\,\delta^{ij}+\left(\Gamma_\mu-iV_\mu^{(s)}\right)\delta^{ij}-i\epsilon^{ijk}\Tr\left(\tau^k\,\Gamma_\mu\right),
\end{eqnarray}
with the isospin indices $i,j,k=(1,2,3)$. As we are only working with two 
light 
flavors $u,\,d$, we identify $V_\mu^{(s)}=\frac{e}{2}\,A_\mu$ as the isoscalar 
component of the external electromagnetic field. The isospin-3/2 
projector is defined as usual as $\xi^{ij}_{I=3/2}=\delta^{ij}-\frac{1}{3}\,
\tau^i\tau^j$. As explained in section \ref{general}, we are explicitly taking into account the finite nucleon-delta mass splitting already at leading order, denoted by the parameter $\Delta$ in Eq.(\ref{eq:lag1}).
Furthermore, 
$g_A$ and $g_1$ are the axial nucleon and delta coupling constants. It turns out that $g_1$ is not needed for the calculation of the magnetic moments of the nucleon to the order considered in this work.

\subsubsection{Next-to-leading order (NLO) Lagrangian}

The less known NLO nucleon and delta Lagrangians can be found in \cite{review}. Here we only discuss the terms pertaining to magnetic moments. Defining 
the chiral field tensors via
\begin{eqnarray}
f_{\mu\nu}^+&=&\sqrt{U}^{\,\dagger}\left\{\partial_\mu F_\nu^R-\partial_\nu F_\mu^R-i\left[F_\mu^R,F_\nu^R\right]\right\}\sqrt{U}
+\sqrt{U}\left\{\partial_\mu F_\nu^L-\partial_\nu F_\mu^L-i\left[F_\mu^L,F_\nu^L\right]\right\}\sqrt{U}^{\,\dagger}\nonumber \\
V_{\mu\nu}^{(s)}&=&\partial_\mu V_\nu^{(s)}-\partial_\nu V_\mu^{(s)}\;,
\label{eq:fmuplus}
\end{eqnarray}
we utilize the following representation of the NLO nucleon and delta Lagrangians \cite{review}
\begin{eqnarray}
{\cal L}^{(2)}_{N}&=&\bar{N}_v\left\{
-\frac{i}{4\,M}\left[S^\mu,S^\nu\right]
\left(\left(1+\kappa_v^0\right)f_{\mu\nu}^++2\left(1+
\kappa_s^0\right)V_{\mu\nu}^{(s)}\right)+\ldots\right\}N_v\; ,\label{eq:kappa}\\
{\cal L}^{(2)}_{\Delta}&=&\bar{T}_i^\mu\left\{-\frac{1}{2 M}\left[[S^\alpha,S^\beta]\left(D^{ik}_\alpha D^{kj}_\beta-D^{ik}_\beta D^{kj}_\alpha\right)g_{\mu\nu}+a_6\,i f_{\mu\nu}^+\delta^{ij}+2 a_7\,i V_{\mu\nu}^{(s)}\delta^{ij}\right]+\dots\right\}\,T^\nu_j\;.\nonumber
\end{eqnarray}
The two couplings $\kappa_v^0$ and $\kappa_s^0$
correspond to the (bare) isovector and isoscalar anomalous magnetic moments of
the nucleon, taken in the chiral limit. Their strength is determined
by physics which lies outside of the chiral effective field theories. We therefore 
treat them as free parameters to be determined by a fit to
lattice simulations of the magnetic moments, as discussed in section \ref{numerik}.
Likewise, $a_6,\,a_7$ are the two corresponding anomalous magnetic dipole moments 
of $\Delta$(1232). It turns out that these two couplings do 
not contribute in the calculation of the magnetic moments of the nucleon to 
leading one-loop order---we therefore relegate a discussion of these interesting quantities to forthcoming work.

\subsection{Modified $N\Delta$ Transition Lagrangian} \label{modified}

While the leading order $\pi N$ and $\pi\Delta$ interaction Lagrangians discussed in 
the previous section follow the standard 
rules of chiral power counting and have been used in many calculations, we now
present a modified version, more appropriate for our purposes,  of the leading order 
chiral nucleon-delta transition Lagrangian in the presence of an external
4-vector electromagnetic background field $A_\mu$. The {\em leading order} 
$N\Delta$ transition Lagrangian we propose has the form
\begin{eqnarray}
{\cal L}^{(1)}_{N\Delta}&=&\bar{T}^\mu_i\left[c_A\,w^i_\mu+
c_V\,i\,f_{\mu\nu}^{+\,i}S^\nu\right]N_v\,+\,h.c.\; ,
\label{eq:cv}
\end{eqnarray}
with 
\begin{eqnarray}
w^i_\mu=\frac{1}{2}\Tr\left(\tau^i u_\mu\right)\,,\quad\quad
f_{\mu\nu}^{+\,i}&=&\frac{1}{2}\Tr\left(\tau^if_{\mu\nu}^+\right)\; .
\end{eqnarray}
It involves the axial (transition) coupling
$c_A$ as well as the (iso-) vector (transition) coupling $c_V$, 
which govern
the strengths of the $\pi N\Delta$ and $\gamma N\Delta$ vertices 
{\em in the chiral limit}, respectively.
In section \ref{numerik} we will discuss numerical estimates for
these parameters. It is standard practice to include the axial $N\Delta$ transition 
in the leading order nucleon-delta transition 
Lagrangian ({\it e.g.} see \cite{review,cohen}) via the coupling $c_A$, whereas the {\em leading term} of (iso)vector $N\Delta$ 
transition is usually only taken into account at sub-leading 
order in the $N\Delta$ transition Lagrangian (see {\it e.g.} Eq.(112) of 
Ref.\cite{review}). The main reason for this asymmetric treatment of the 
axial and vector $N\Delta$ transition lies in the ``standard'' counting
(``naive power counting'') of the chiral tensors in powers of the generic
mass or momentum scale $\epsilon$: The (pseudo-) vector $w_\mu^i$ scales as order $\epsilon$,
whereas $f_{\mu\nu}^+$ has dimension $\epsilon^2$. We note that these counting rules on the one hand lead to an asymmetric treatment of the vector, axial-vector $N\Delta$-transitions by distributing them onto the NLO, respectively LO Lagrangians. On the other hand they result in a symmetric (NLO) treatment among the {\em magnetic} $\gamma NN,\,\gamma N\Delta$ and $\gamma\Delta\Delta$ couplings, which, for example, is appealing from the viewpoint of the SU(6) quark model. 

In this work we propose the ansatz---displayed in Eq.(\ref{eq:cv})---to 
promote the ``off-diagonal'' magnetic $N\Delta$ transition into the leading order
$N\Delta$ transition Lagrangian, while leaving the corresponding ``diagonal''
$\gamma NN,\,\gamma\Delta\Delta$ couplings $\kappa_v^0,\,a_6$ 
in the NLO Lagrangians of Eq.(\ref{eq:kappa}) as suggested by
dimensional analysis\footnote{One may wonder why we do not propose to promote all three couplings to the respective leading order Lagrangians. We refrain from doing so because for the ``diagonal'' couplings $\kappa_v^0,\,a_6$ this would lead to the peculiar situation that the anomalous (``Pauli'') contribution to the nucleon, delta magnetic moments would come in at leading order, while the 1/M suppressed regular (``Dirac'') contribution is generated in subleading order via the non-relativistic reduction.}. In order visualize that the isovector transition structure in Eq.(\ref{eq:cv}) is part of the leading order Lagrangian, we assign an intrinsic power $\epsilon^{-1}$ to it 
\begin{eqnarray}
c_V\equiv c_V^{(-1)} \; ,
\end{eqnarray}
rendering the structure $c_V^{(-1)}\,f_{\mu\nu}^+$ to scale as ${\cal O}(\epsilon)$, as expected for a leading order Lagrangian. In Ref.\cite{Nstar} it was argued that this intrinsic scaling of the (dimensionful) coupling $c_V$ should be made more explicit by introducing a dimension-free coupling $\tilde{c}_V$ via $c_V^{(-1)}\equiv\tilde{c}_V/\Delta$. On a computational level that prescription is of course equivalent to the structure in Eq.(\ref{eq:cv}). However, given that we are after the quark mass dependence of the magnetic moments, we do not want to prejudice as to which dimensionful quantity---be it $\Delta$ or $M_N$---sets the scale in $c_V^{(-1)}$ to result in a dimension-free coupling $\tilde{c}_V$. The reason for this caution lies in the fact that the two mass scales $\Delta$ and $M_N$ have quite a different intrinsic quark mass dependence of their own (see {\it e.g.} the lattice simulations given in Ref.\cite{latticemasses}), which would then (in higher orders of the calculation) seem to alter the chiral extrapolation curve\footnote{The counterterms in the chiral effective field theory do of course not depend on the quark masses. However, the problem discussed here arises if one makes the identification between the associated chiral limit parameters ({\it e.g.} the ``bare'' nucleon mass $M_N^0$) and physical quantities (like $M_N=0.938$ MeV).} depending on which parameterization was used for the leading isovector $N\Delta$-transition Lagrangian. In order to avoid this ambiguity we therefore choose to work with a dimensionful coupling $c_V\equiv c_V^{(-1)}$, keeping in mind that it carries an intrinsic power of $\epsilon^{-1}$.

One of the motivations given in Ref.\cite{Nstar} that the magnetic $N\Delta$ transition coupling should scale as $\epsilon^{-1}$, whereas the magnetic $NN,\,\Delta\Delta$ couplings of Eq.(\ref{eq:fmuplus}) obey the standard chiral counting rules $\sim\epsilon^0$, was the well known fact that quark model SU(6) symmetry factors relating $c_V$ to the isovector anomalous moment of the nucleon typically underestimate the $\gamma N\Delta$ transition strength by as much as 30\% \cite{SU6}. This observation and the fact that many observables in pion photo-/electroproduction \cite{piphoto} as well as in nucleon Compton scattering \cite{Compton} crucially depend on a proper treatment of the M1 $\gamma N\Delta$ transition in the chiral effective field theory provide a {\it physical} motivation for a more prominent role of this important structure, aside from the formal discussion given above. 

However, the main reason why we insist on having the magnetic $N\Delta$-transition to be part of the leading order Lagrangian Eq.(\ref{eq:cv}) does not lie in its mere numerical strength. After all the coupling $c_V$---which  only represents the leading term of this transition, taken in the chiral limit---might be substantially different\footnote{The chiral expansion of the M1 $\gamma N\Delta$ transition form factor based on ``naive power counting'' is analyzed in Refs.\cite{bss,ndelta}. Given that there are three additional counterterms \cite{ndelta} contributing to this transition, we do not have good information on the strength of the coupling $c_V$ at the moment.} from the physical M1 $\gamma N\Delta$ transition strength. Even more important from the point of view of the quark mass expansion of the magnetic moments is the fact that this operator---as will be discussed in sections \ref{calculation},\ref{results}---produces important non-analytic quark mass dependence in the magnetic moments, which turns out to be essential for a meaningful chiral extrapolation. Nevertheless one cannot proceed at will promoting arbitrary operators into any order of the Lagrangian. One has to show {\em explicitly} that the resulting effective field theory can still be renormalized. For the specific calculation considered here we demonstrate in the course of this paper that this is indeed the case.

To summarize this central paragraph of our work: As a result of our proposal Eq.(\ref{eq:cv}) one obtains a symmetric treatment between the axial 
and the (iso)vector $N\Delta$ transitions and, accordingly, a modified diagrammatic expansion. The consequences of this procedure for the case 
of the chiral expansion of the anomalous magnetic 
moments of the nucleon are the main physics topic of this paper\footnote{Obviously there are interesting applications of this proposal for calculations of electromagnetic scattering processes \cite{piphoto,Compton} in effective field theory which are being explored \cite{consequences}.}. The chiral
(i.e. quark-mass) expansion of any observable is of course not changed by 
boosting operators into different orders of the effective Lagrangian. However, different expansion schemes can bring in important operators 
already at lower orders in the calculation, thus making a differing scheme more effective. Before we can go into the calculation, however, we first have to specify some higher order couplings required for
a systematic calculations with our modified power-counting.

\subsection{N$^2$LO Nucleon Lagrangian}\label{n2lo}

The standard ChPT calculation of the magnetic moments in the heavy baryon limit gives a finite result at leading one-loop order without any counterterm \cite{TW}. Allowing for the possibility that the pion-cloud can also fluctuate around an intermediate spin 3/2 baryon requires the introduction of two counterterms proportional to the octet-decuplet mass splitting $\Delta$ to be able to renormalize the leading
one-loop diagrams in a theory with explicit pion, nucleon and delta degrees of freedom based on ``naive power counting'', as was shown in Ref.\cite{BFHM}. Labeling the two corresponding structures $D_1$ and $D_2$, we note that they are related to the couplings $B_{28},\, B_{29}$ of Ref.\cite{BFHM} via $D_1=B_{28}/(4\pi\fpi)^2,\,D_2=B_{29}/(4\pi\fpi)^2$. Here we again prefer the notation involving dimensionful couplings in order to avoid speculating about the underlying mass scale and its inherent quark mass dependence, analogous to our reasoning regarding $c_V$ in section \ref{modified}. As explained in Ref.\cite{BFHM} the introduction of these two couplings is needed for the renormalization of the magnetic $\gamma NN$ vertex function but does not lead to observable consequences, as the two structures are quark mass independent. The finite parts of these couplings can therefore be utilized to guarantee decoupling of the delta resonance in the limit of fixed quark masses and $\Delta\rightarrow\infty$ for any value of the 
regularization scale $\lambda$, as will be discussed in section \ref{calculation}.

If one now moves on and modifies the ``naive power counting''---as we propose in Eq.(\ref{eq:cv})---one is not surprised to learn that this also leads to consequences in the {\em most general} N$^2$LO nucleon counterterm Lagrangian needed to renormalize the leading-one-loop diagrams. For the particular case of our magnetic moment calculations we find---to leading one-loop order---that moving the magnetic $N\Delta$ transition into the leading order Lagrangian as proposed in Eq.(\ref{eq:cv}) in general induces four N$^3$LO operators with coupling constants $D_3,\,D_4,\,E_1,\,E_2$ to move down into the N$^2$LO Lagrangian\footnote{In order to construct a complete set of N$^2$LO nucleon counterterms required to renormalize all possible 1-loop graphs 
involving pions, nucleons and deltas in the presence of arbitrary external fields for the here proposed new form of the leading order $N\Delta$ transition Lagrangian Eq.(\ref{eq:cv}), one would have to perform a new one-loop renormalization analysis for single nucleon processes as for example done in Ref.\cite{Ecker}, which is 
beyond the scope of this article.}:
\begin{eqnarray}
{\cal L}^{(3)}_{N}&=&\bar{N}_v\left\{D_1\,\Delta\,i\left[S^\mu,S^\nu\right]f_{\mu\nu}^++D_2\,\Delta\,i\left[S^\mu,S^\nu\right]V_{\mu\nu}^{(s)}\right.\nonumber\\
& &+D_3\,\Delta^2\,i\left[S^\mu,S^\nu\right]f_{\mu\nu}^++D_4\,\Delta^2\,i\left[S^\mu,S^\nu\right]V_{\mu\nu}^{(s)}\nonumber\\                      
& &+E_1\,i\left[S^\mu,S^\nu\right]\chi_+^{(s)}\,f_{\mu\nu}^++E_2\,i\left[S^\mu,S^\nu\right]\chi_+^{(s)}\,V_{\mu\nu}^{(s)}\nonumber\\
& &\left.
-\frac{1}{8
M^2}\left[[S^\mu,S^\nu]\left((1+2\kappa_v^0)\,f_{\mu\sigma}^++2(1+2\kappa_s^0)\,V_{\mu\sigma}^{(s)}\right)v^\sigma
D_\nu+h.c.\right]+\ldots\right\}N_v\; ,\nonumber\\
& & \label{eq:e1e2}
\end{eqnarray}
with
\begin{eqnarray}
\chi_+^{(s)}&=&\frac{1}{2}\Tr\left(\sqrt{U}^{\,\dagger}\chi
\sqrt{U}^{\,\dagger}+\sqrt{U}\chi^\dagger \sqrt{U}\right) .
\end{eqnarray}
Note that in the following we work in the isospin limit $m_u=m_d=\hat{m}$ and therefore only need to consider
the isoscalar component of $\chi_+$. Throughout this
calculation we utilize dimensional regularization and denote the resulting
infinities by the quantity $L$ spelled out in Appendix \ref{A}. All six counterterms then have the generic structure
\begin{eqnarray}
{\cal L}_N^{c.t.}=C_i\,\bar{N}_v \,O_i^{(3)}\,N_v\,,
\end{eqnarray}
with $C_i=C_i^r(\lambda)+\beta_i\,16\pi^2 L$ and 
the associated $\beta-$functions given in Table \ref{beta}. We 
observe that 
the four counterterms can be separated
into a scale ($\lambda$) dependent finite and an infinite part. 
\begin{table}
\begin{tabular}{|c|c|c|}\hline
LEC & $O_i$ & $\beta_i$ \\\hline
$D_1$ & $\Delta \,i [S^\mu,S^\nu] \,f_{\mu\nu}^+$ & $+c_A^2/\left(18\,\pi^2\fpi^2\right)$\\
$D_2$ & $\Delta \,i [S^\mu,S^\nu] \,V_{\mu\nu}^{(s)}$ & -- \\
$D_3$ & $\Delta^2 \,i [S^\mu,S^\nu] \,f_{\mu\nu}^+$ & $-c_Ac_Vg_A/\left(27\,\pi^2\fpi^2\right)$ \\
$D_4$ & $\Delta^2 \,i [S^\mu,S^\nu] \,V_{\mu\nu}^{(s)}$ & -- \\
$E_1$ & $i [S^\mu,S^\nu] \,\chi_+^{(s)} \,f_{\mu\nu}^+$ & $+c_Ac_Vg_A/\left(36\,\pi^2\fpi^2\right)$\\
$E_2$ & $i [S^\mu,S^\nu] \,\chi_+^{(s)} \,V_{\mu\nu}^{(s)}$ & -- \\\hline
\end{tabular}
\caption{N$^2$LO counterterms and their $\beta$-functions contributing to the magnetic moments of the nucleon to leading one-loop order.\label{beta}}
\end{table}
Once more we want to stress that based on ``naive powercounting'' one would not
expect to find the local operators associated with $D_3,\,D_4,\,E_1,\,E_2$ among the terms
of the $N^2$LO Lagrangian, as their structures $\chi_+^{(s)}$ and $f_{\mu\nu}^+$, respectively $V_{\mu\nu}^{(s)}$ scale as $\sim\epsilon^2$ each in standard counting. In contrast to the coupling $c_V$---which, in the
previous section, has been attributed an intrinsic power $\epsilon^{-1}$ due to
its importance---$D_3,\,D_4,\,E_1$ and $E_2$ obtain their intrinsic power of
$\epsilon^{-1}$ solely by the requirement that every effective field theory
based on the most general chiral Lagrangian should be renormalizable, 
independent of
the particular organization of the perturbative expansion. Following the notation of section \ref{modified} we therefore write
\begin{eqnarray}
E_1\equiv E_1^{(-1)}\;;&\quad\quad& E_2\equiv E_2^{(-1)}\;, \nonumber\\
D_3\equiv D_3^{(-1)}\;;&\quad\quad& D_4\equiv D_4^{(-1)}\;, 
\end{eqnarray}
rendering the structures $\left[D_3^{(-1)}\,\Delta^2\,f_{\mu\nu}^+\right]$, $\left[D_4^{(-1)}\,\Delta^2\,V_{\mu\nu}^{(s)}\right]$, $\left[E_1^{(-1)}\,\chi_+^{(s)}\,f_{\mu\nu}^+\right]$ and $\left[E_2^{(-1)}\,\chi_+^{(s)} \,V_{\mu\nu}^{(s)}\right]$ to scale as $\sim\epsilon^3$, in accordance with the power of the N$^2$LO nucleon Lagrangian.

Finally, we comment on the observation that three of the six $\beta-$functions
in Table \ref{beta} are zero, suggesting that three counterterms would be
sufficient to renormalize the magnetic moments to leading one-loop
order. We note, however, that this simplification only occurs for our special choice of representation of the chiral field strength tensors given
in Eq.(\ref{eq:fmuplus}). In the (more widely used) conventions of 
Ref.\cite{BKM}---which encode the chiral field strength tensors as $F_{\mu\nu}^+\equiv f_{\mu\nu}^++2V_{\mu\nu}^{(s)}$ and $\Tr ( F_{\mu\nu}^+)\equiv 4 V_{\mu\nu}^{(s)}$---one would require six counterterms with non-zero $\beta-$functions. Another reason for our retaining of $D_2,\,D_4,\,E_2$ in the N$^2$LO Lagrangian Eq.(\ref{eq:e1e2}) lies in the fact that we want to treat the (unknown) short distance physics\footnote{It also turns out that---for the case of 2 flavors considered here---$E_2$ provides the leading quark mass dependence in the isoscalar sector, see section \ref{isoscalar}.} in the isoscalar and the isovector sector of the theory in a symmetric fashion. With these remarks we close our discussion on the required 
chiral Lagrangians and move on to the details of the calculation.  

\section{The Calculation}\label{calculation}
\setcounter{equation}{0}

At first glance the reader may wonder why we present yet another calculation for the baryons' magnetic moments, as this topic is probably the best studied one in the field of chiral effective field theories. The leading non-analytic quark mass dependence is known since the 1970s \cite{Pagels}, calculations with (octet) meson and (octet) baryon degrees of freedom were pioneered in baryon ChPT in the late 1980s \cite{Gasser}, the advent of heavy baryon techniques brought about many more calculations throughout the 1990s, including the first studies with intermediate decuplet baryons \cite{Jenkins}. An overview of some recent calculations and references can be found in \cite{revisited}.

When we present our leading-one-loop results for the anomalous magnetic moments of the nucleon arising from nucleon, delta and pion degrees of freedom as given by the diagrams in Fig.\ref{fig:diags} and the corresponding amplitudes in Appendix \ref{A}, we do {\em not} claim that we have calculated any new contributions previously not considered in the literature. In fact, Ref.\cite{Jenkins} contains even more one-loop diagrams than we consider here. We point out that the underlying philosophy between our work and for example Ref.\cite{Jenkins} is a different one. In \cite{Jenkins} the authors have calculated all possible one-loop topologies contributing to the magnetic moments based on the {\em leading} photon and meson couplings to octet/decuplet baryons. In this work our powercounting---as discussed in section \ref{Lagrangians}---establishes a hierarchy among the one-loop diagrams, selecting the ones given in Eq.(\ref{amplitudes}) to be leading-one-loop ($\equiv {\cal O}(\epsilon^3)$) and dictates the structure of counterterms to be included at this order (c.f. Eq.(\ref{eq:e1e2})). Given that we spent the first part of this paper arguing for a ``modified powercounting'' it should be obvious that chiral effective field theory for low energy baryon properties does not possess one unique perturbative expansion parameter like $\alpha_{QED}$ in Quantum Electrodynamics. Due to the complex structure of the low energy hadron spectrum and the many different scales involved several counting approaches have to be explored, our proposal in Eq.(\ref{eq:cv}) is only one possibility, albeit a well-motivated one. Ultimately the success in describing phenomena determines which expansion scheme really is ``effective''. With this in mind we point out that our leading-one-loop calculation of the anomalous magnetic moments of the nucleon depends on only a few structures and parameters, providing some ``predictive'' power with respect to the few lattice data available at the moment. This will be discussed in the upcoming subsections. Of course it is possible to push the calculation to one higher order ($\equiv {\cal O}(\epsilon^4)$) \cite{massimiliano}, which formally includes 35 extra diagrams, containing the remaining ones of Ref.\cite{Jenkins} as well as some new ones\footnote{At ${\cal O}(\epsilon^4)$ one also has to take into account the first 1/M corrections to the axial $NN$ and $N\Delta$ vertices, {\it e.g.} see \cite{review}.}. On the other hand, at ${\cal O}(\epsilon^4)$ one encounters at least four additional unknown couplings, making the comparison to lattice data more strenuous, so we defer it to a later stage \cite{massimiliano}.

The NLO calculation discussed here proceeds in a straightforward manner, the results can be found in Appendix \ref{A}. To the order we are working 10 one loop topologies displayed in Fig.\ref{fig:diags} 
have to be analyzed---of which only 4 yield non-zero results (c.f. Appendix \ref{A}). We note that the isovector $N\Delta$-transition coupling $c_V$ discussed in the previous section contributes to two of the loop diagrams. We utilize dimensional regularization\footnote{For recent work analyzing the magnetic moments of the octet baryons employing lattice regularization methods see Ref.\cite{Randy}.} throughout, the $\beta$-functions of the six counterterms are given in Table \ref{beta}. The finite parts of the counterterms $E_1^r(\lambda),\,E_2^r(\lambda)$ will be left as free parameters and fixed from lattice data in section \ref{numerik}. However, as already indicated in section \ref{n2lo}, matters are different in the case of $D_i^r,\,i=1,\dots 4$:
The infinite parts of these four counterterms are utilized to cancel divergences $\sim\Delta L$ and $\sim\Delta^2L$ (c.f. Table \ref{beta}). The finite parts of these four structures cannot be observed separately from the chiral limit structures $\kappa_v^0,\,\kappa_s^0$. As suggested in Ref.\cite{BFHM}, one can make use of this freedom in implementing decoupling of the Delta resonance, {\it e.g} we demand that the theory with explicit delta degrees of freedom transforms itself into a theory with just pion and nucleon degrees of freedom in the limit $\Delta\rightarrow\infty$ (for fixed quark masses). This implies, for example, that all quark mass independent polynomial structures in $\Delta$ must vanish, for any value of the chosen regularization scale $\lambda$. With the results from Appendix A we find that the assignment
\begin{eqnarray}
D_1^r(\lambda)&=&\frac{c_A^2}{36\pi^2\fpi^2}\left\{2\log\left(\frac{2\Delta}{\lambda}\right)-\frac{5}{3}\right\}\;,\nonumber\\
D_2^r(\lambda)&=&0\;,\nonumber\\
D_3^r(\lambda)&=&\frac{c_Ac_Vg_A}{27\pi^2\fpi^2}\left\{\frac{1}{6}-\log\frac{2\Delta}{\lambda}\right\}\;, \nonumber\\
D_4^r(\lambda)&=&0\;,\label{eq:decoupling}
\end{eqnarray}
satisfies the condition of vanishing polynomial structures. The decoupling of the Delta resonance achieved by this choice can be best seen in the limit $m_\pi/\Delta\rightarrow 0$, which we discuss in section \ref{chiral}. After these technical comments we finally proceed to the physical results.

\begin{figure}[!htb]
  \begin{center}
    \includegraphics*[width=0.9\textwidth]{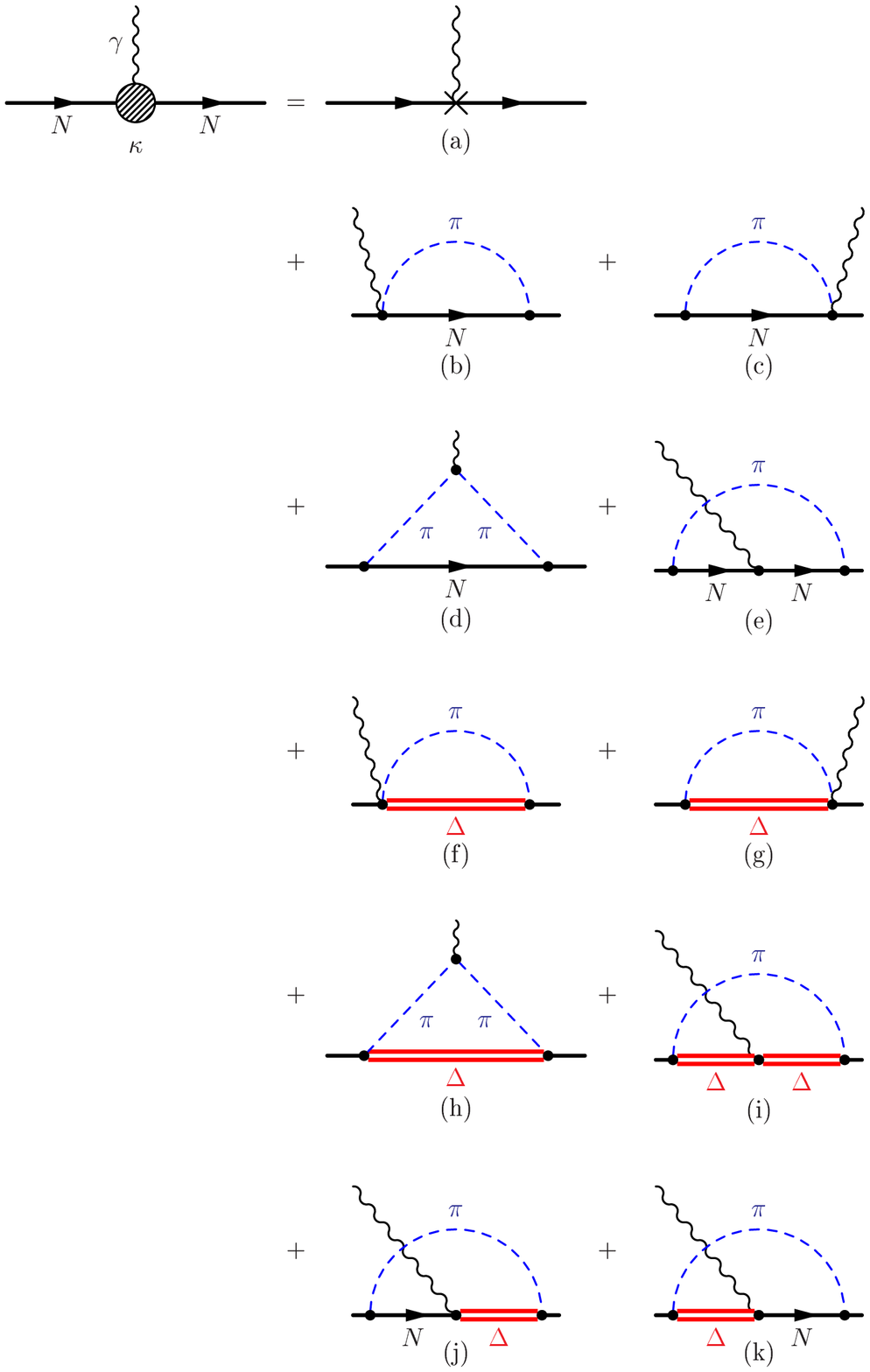}
    \caption{Diagrams contributing to the anomalous magnetic moment of the nucleon at leading-one-loop order.}
    \label{fig:diags}
  \end{center}
\end{figure}


\section{Analytic Results}\label{results}
\setcounter{equation}{0}


\subsection{General Remarks}
We now turn to the quark mass dependence of the proton and neutron magnetic
moments, $\mu_p$ and $\mu_n$. Two remarks are in order at this point. First, we 
note that the chiral corrections concern only the anomalous parts $\kappa_p,\;\kappa_n$ of the magnetic moments 
\begin{eqnarray}
\mu_p=1+\kappa_p\;, & &\mu_n=\kappa_n\;.
\end{eqnarray}
Second, the chiral corrections affect the isovector and isoscalar anomalous magnetic moments of the nucleon $\kappa_v,\,\kappa_s$
quite differently. We therefore discuss our results in the isospin basis defined by
\begin{eqnarray}
\kappa_v=\kappa_p-\kappa_n\,,\; & & \kappa_s=\kappa_p+\kappa_n\;.
\end{eqnarray}
Both $\kappa_v$ and $\kappa_s$ are functions of the isospin averaged quark mass\footnote{In addition to neglecting all effects
from strong isospin breaking---i.e. the $m_u-m_d$ mass difference---we also do
not consider electromagnetic corrections arising from the different charges of
the light $u$ and $d$ quarks. All other quarks taken to be infinitely heavy and
are effectively integrated out of the theory.} $\hat{m}=(m_u+m_d)/2$ and of the
chiral condensate parameter $B_0$. These two quantities are combined to form the leading order term in the quark mass 
expansion of the (squared) pion mass. One finds \cite{GL}
\begin{eqnarray}
m_\pi^2&=&2\hat{m}B_0\left\{1+O(\hat{m}B_0)\right\} ,
\end{eqnarray}
which to leading order corresponds to the well-known Gell-Mann, Oakes, Renner relation \cite{GOR}.
Changing the value of the light quark mass $\hat{m}$ (in a numerical simulation
of QCD) therefore leads to a quadratic change of the pion mass (modulo higher order corrections). 
In the following we present our results as functions
of  $m_\pi$. When comparing a magnetic moment calculation performed in a chiral effective field theory to lattice QCD data, one therefore requires these data as a function of the mass of the lowest lying $0^-$ boson in the simulation, which is identified as the corresponding lattice pion. It is understood, that both the pion mass and the associated nucleon magnetic moment simulation are performed with identical lattice parameter input. Such correlated lattice results have been reported in the literature \cite{adelaide} and we will discuss them in section \ref{numerik}. Here we focus on the analytic results.

\subsection{Isovector Anomalous Magnetic Moment}

\subsubsection{Analytic Results to leading one-loop order}

All results presented in this section can be directly read off from the amplitudes shown in Appendix \ref{A}, refering to the relevant Feynman diagrams shown in Fig.\ref{fig:diags}. 
.
The NLO result of SU(2) non-relativistic (``heavy'') baryon ChPT with only pion and nucleon degrees of freedom (in the following denoted as case $A$) reproduces the well-known \cite{Pagels}
 leading non-analytic quark mass correction to the isovector anomalous magnetic moment of the nucleon
\begin{eqnarray}
\kappa_v^A&=&\kappa_{v,\,A}^0-\frac{g_A^2\,m_\pi M}{4\pi \fpi^2}+N^2LO \;.\label{sa}
\end{eqnarray}
The standard power-counting of heavy baryon ChPT tells us that to NLO this result originates from graphs a) ... e) of Fig.\ref{fig:diags}. All other contributions are relegated to 
higher order corrections\footnote{The same calculation performed to NLO in relativistic baryon ChPT is discussed in Appendix \ref{B}. It contains many more structures which are part of the higher order corrections in the heavy baryon approach discussed here.}.

Recently the chiral corrections to $\kappa_v$
have also been evaluated to NLO in the SU(2) Small Scale Expansion approach \cite{BFHM}, which includes explicit pion, nucleon and delta degrees of freedom
(in the following denoted as case $B$). With the additional counting prescription for the finite scale $\Delta$ (c.f. section \ref{general}), this calculation follows the standard (``naive'') power counting rules of baryon ChPT and thus only includes the coupling $c_A$ in the $N\Delta$ transition Lagrangian of Eq.(\ref{eq:cv}): 
\begin{eqnarray}
\kappa_v^B&=&\kappa_{v,\,B}^0-\frac{g_A^2\,m_\pi M}{4\pi \fpi^2} \nonumber\\
          & &+\frac{2 c_A^2 \Delta M}{9\pi^2\fpi^2}
             \left\{\sqrt{1-\frac{m_\pi^2}{\Delta^2}}\log
             \left[R(m_\pi)\right]+\log\left[\frac{m_\pi}{2\Delta}\right]
                              \right\}+N^2LO \,,\label{sb}
\end{eqnarray}
with
\begin{eqnarray}
R(m_\pi)&=&\frac{\Delta}{m_\pi}+\sqrt{\frac{\Delta^2}{m_\pi^2}-1}\;.
\end{eqnarray}
To NLO this result originates from graphs a) ... i) of Fig.\ref{fig:diags}, all other contributions are again relegated to higher orders.

Now we present the chiral corrections to
$\kappa_v$ calculated to NLO in our modified scheme 
(in the following denoted as case $C$).
In contrast to the calculation of Ref.\cite{BFHM}, we use the modified leading order nucleon-delta transition Lagrangian  
Eq.(\ref{eq:cv}) which includes $c_V$---the leading term of the isovector $\gamma N\Delta$ M1 transition---as well as the induced additional N$^2$LO counterterm $E_1$ of Eq.(\ref{eq:e1e2}),
required for renormalization of the extra one-loop graphs involving $c_V$. We obtain
\begin{eqnarray}
\kappa_v^C&=&\kappa_{v,\,C}^0-\frac{g_A^2\,m_\pi M}{4\pi \fpi^2}\nonumber\\
          & &               +   \frac{2 c_A^2 \Delta M}{9\pi^2\fpi^2}
                              \left\{\sqrt{1-\frac{m_\pi^2}{\Delta^2}}\log\left[R(m_\pi)\right]+\log\left[\frac{m_\pi}{2\Delta}\right]
                              \right\} \nonumber\\
          & &               -   8 E_1(\lambda) Mm_\pi^2
                         +  \frac{4c_A c_V g_A M m_\pi^2}{9\pi^2\fpi^2}\log\left[\frac{2\Delta}{\lambda}\right] \nonumber \\
          & &            +  \frac{4c_A c_V g_A M m_\pi^3}{27\pi\fpi^2\Delta}
                            \nonumber\\
          & &            -   \frac{8 c_A c_V g_A \Delta^2 M}{27\pi^2\fpi^2}
                              \left\{\left(1-\frac{m_\pi^2}{\Delta^2}\right)^{3/2}\log\left[R(m_\pi)\right]+\left(1-\frac{3m_\pi^2}{2\Delta^2}\right)
                              \log\left[\frac{m_\pi}{2\Delta}\right]
                              \right\} \nonumber\\
          & &            +N^2LO \;.\label{sc}
\end{eqnarray}
This NLO result arises from the graphs displayed in Fig.\ref{fig:diags}. We note that very few diagrams, of which only 5 are non-zero (c.f. Appendix \ref{A}) to this order, produce such an intricate quark mass dependence in $\kappa_v$, making the modified scheme proposed here very effective
in calculating chiral corrections for electromagnetic quantities. 
Before we study the quantitative differences in the chiral corrections to $\kappa_v$
between schemes $A,B$ and $C$, we first draw
a qualitative picture of the extra physics contained in scheme $C$. For a proper comparison we therefore move on to a discussion concerning the chiral limit.

\subsubsection{Chiral Limit Results} \label{chiral}

For completeness we show here the chiral limit results of the three different 
effective field theory calculations for $\kappa_v$ 
discussed above---all calculated to NLO accuracy by Taylor expansions of Eqs.(\ref{sa},\ref{sb},\ref{sc}):
\begin{eqnarray}
\kappa_v^A|_{NLO}&=&\kappa_{v,\,A}^0-\frac{g_A^2 M}{4\pi \fpi^2}\,m_\pi\nonumber\\
\kappa_v^B|_{NLO}&\approx&\kappa_{v,\,B}^0-\frac{g_A^2 M}{4\pi \fpi^2}\,m_\pi
                   +m_\pi^2\left[-\frac{c_A^2 M}{18\pi^2\fpi^2\Delta}
                   +\frac{c_A^2 M}{9\pi^2\fpi^2\Delta}\log\left(
                   \frac{m_\pi}{2\Delta}\right)+N^2LO\right]\nonumber\\
                & &+m_\pi^3\left[0+N^2LO\right]
                   +m_\pi^4\left[\frac{c_A^2 M}{144\pi^2\fpi^2\Delta^3}
                   +\frac{c_A^2 M}{36\pi^2\fpi^2\Delta^3}
                   \log\left(\frac{m_\pi}{2\Delta}\right)
                   +N^2LO\right]+\ldots\nonumber\\
\kappa_v^C|_{NLO}&\approx&\kappa_{v,\,C}^0
                  -\frac{g_A^2 M}{4\pi \fpi^2}\,m_\pi\nonumber\\
& &+m_\pi^2\left[-\frac{c_A^2 M}{18\pi^2\fpi^2\Delta}
                   +\frac{c_A^2 M}{9\pi^2\fpi^2\Delta}\log\left(
                   \frac{m_\pi}{2\Delta}\right)\right.\nonumber\\
& &\phantom{+m_\pi^2}\left.
                   -8 E_1(\lambda) M+\frac{4c_A c_V g_A M}{9\pi^2\fpi^2}
                   \log\left(\frac{2\Delta}{\lambda}\right)
                   +\frac{2c_Ac_Vg_AM}{27\pi^2\fpi^2}
                   +N^2LO\right]\nonumber\\
                & &+m_\pi^3\left[\frac{4c_A c_V g_A M}{27\pi\fpi^2\Delta}
                   +N^2LO\right]\nonumber\\
                & &+m_\pi^4\left[\frac{c_A^2 M}{144\pi^2\fpi^2\Delta^3}
                   +\frac{c_A^2 M}{36\pi^2\fpi^2\Delta^3}
                   \log\left(\frac{m_\pi}{2\Delta}\right)
                   -\frac{c_Ac_Vg_AM}{12\Delta^2\pi^2\fpi^2}
                   +\frac{c_Ac_Vg_AM}{9\Delta^2\pi^2\fpi^2}
                   \log\left(\frac{m_\pi}{2\Delta}\right)\right.\nonumber\\
& &\phantom{+m_\pi^4}\left.+N^2LO\right]\nonumber\\
& &+\ldots\label{eq:chiral}
\end{eqnarray}
First we observe that schemes B and C produce a whole string of terms proportional to $m_\pi^n$. We note that all these terms---with exception of the chiral limit couplings $\kappa_v^0$---do occur {\em at the same NLO order} in the chiral powercounting. They just represent the first few terms of an infinite Taylor series in $m_\pi$ arising from the chiral limit expansion of the logarithms in Eqs.(\ref{sb},\ref{sc}). We note that {\em all} these Taylor coefficients starting from $m_\pi^2$ will receive corrections at N$^2$LO and higher orders. One also observes that the decoupling of the Delta resonance as discussed in section \ref{calculation} is manifest\footnote{The logarithmic scale dependence $\sim\log\Delta/\lambda$ in Eq.(\ref{eq:chiral}) retains the information about the existence of the delta resonance in the decoupling limit $\Delta\rightarrow\infty$. Decoupling is presumably already achieved for finite $\Delta$ if $\Delta\geq\Lambda_\chi$, where $\Lambda_\chi$ denotes the chiral symmetry breaking scale.} in Eq.(\ref{eq:chiral}), based on the counterterm prescription given in Eq.(\ref{eq:decoupling}). 

One can now clearly see that the power counting of scheme $C$ incorporates the
first four terms in the chiral expansion of $\kappa_v$ already at
NLO. Specifically, we note that the structures proportional to 
$m_\pi^3$ are absent to this order\footnote{Structures $\sim m_\pi^3$ are also generated at NLO in relativistic Baryon ChPT, as discussed in Appendix \ref{B}.} in schemes $A$ and $B$. In the traditional expansion scheme $A$
of Heavy Baryon ChPT such terms could only be generated at $N^3LO$ (i.e. at the two-loop level), thus making the explicit calculation of such terms extremely prohibitive. The analytic results presented here and in the previous section are completely general. In order to produce chiral extrapolation functions for the magnetic moments which connect lattice data to the physical worldline and even to the chiral limit we now move on into the numerical analysis of the results presented so far.


\section{Numerical Results}\label{numerik}
\setcounter{equation}{0}


\subsection{General Remarks}

In principle all couplings and masses---aside from $m_\pi$---occuring in Eqs.(\ref{sa},\ref{sb},\ref{sc}) are to be taken at their values in the chiral limit. However, for most of them the chiral limit values are only poorly known. On the other hand, the difference between the values taken at $m_\pi=0$ and the physical values of the couplings is of higher order, allowing us to resort to physical parameters---given in Table \ref{table1}---in those cases where we have only limited information. 

We note that to the order we are working the nucleon mass $M$ is not really a parameter occuring in the calculation, as can be easily seen from Eq.(\ref{amplitudes}). Accordingly, an overall $M$ can be factored out on the right hand side of Eqs.(\ref{sa},\ref{sb},\ref{sc}). The scale $M$ is just a convention to obtain magnetic moments in units of nuclear magnetons ([n.m.]) \cite{PDG}. For the axial coupling constant $g_A$ of the nucleon we use its physical value \cite{PDG}. Not much is known about its value in the chiral limit $g_A^0$, but recent lattice data suggest that the quark mass dependence of this quantity is rather flat \cite{Schierholz}. For the pion decay constant we utilize its physical value \cite{TW}, as the difference to $f_\pi^0$ is known to be be only a few percent \cite{GL}. In order to fix the parameter $\Delta$---representing the nucleon-delta mass splitting---we employ the dispersion theoretical analysis of Ref.\cite{HDT} to obtain the real part of the complex delta mass $M_\Delta=(1211-i\,50)$ MeV. Recent lattice simulations discussed in \cite{latticemasses} suggest that $\Delta$ also has a rather weak quark mass dependence. Finally, we fix the leading axial $N\Delta$ coupling constant $c_A$ by reproducing the imaginary part\footnote{The value of $c_A$ given in Table \ref{table1} corresponds to a strong decay width of $\Gamma_\Delta=100$ MeV \cite{HDT}. We note that the delta properties given in Ref.\cite{HDT} are evaluated at the T-matrix pole.} of the delta mass given in Ref.\cite{HDT}. Unfortunately the quark mass dependence of the axial $N\Delta$ couplings is not known. We therefore {\em assume} that the physical value for $c_A$ constitutes a decent approximation for $c_A^0$. 

If it turns out that some aspects of the reasoning presented here do not hold, then the other couplings---which are directly fitted to the lattice data in the upcoming two sections---have to compensate for any wrong assignments. We consider this issue to be only a temporary problem. As soon as more low mass lattice data for a variety of nucleon structure operators become available, one can fit all parameters directly to lattice data.

\begin{table}
\begin{tabular}{|l|l|}\hline
Parameter & Physical Value \\\hline\hline
$g_A$ & 1.267 \\\hline
$c_A$ & 1.125 \\\hline
$\fpi$ & 0.0924 GeV \\\hline
$M$ & 0.9389 GeV \\\hline
$\Delta\equiv Re\left[M_\Delta\right]-M$ & 0.2711 GeV \\\hline
\end{tabular}
\caption{Input parameters used in this work. With ``physical value'' we denote
their magnitudes at the point where the lowest lying Goldstone boson in the
theory has the mass of 138 MeV and is identified with the pion. The physical
meaning of these parameters is explained in section \ref{Lagrangians}. The 
nucleon
and pion masses are taken as isospin-averaged. \label{table1}}
\end{table}

\subsection{Numerical Analysis of Schemes A and B}

\begin{figure}[!htb]
  \begin{center}
    \includegraphics*[width=0.9\textwidth]{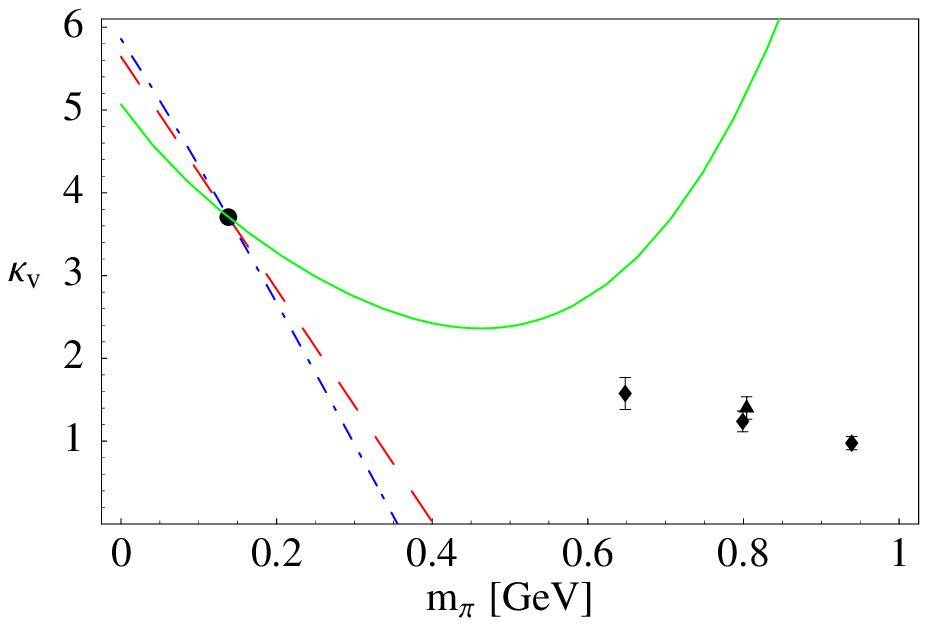}
    \caption{Pion mass dependence of the isovector anomalous magnetic moment in nuclear magnetons. The curves shown denote the 
 standard NLO Heavy Baryon ChPT (dashed line, scheme A, see Eq.(\ref{sa})) and the NLO Small Scale Expansion calculation (dot-dashed line, scheme B, see Eq.(\ref{sb})). The solid line denotes the LO Relativistic BChPT result discussed in Appendix \ref{B}. The lattice data are taken from Ref.\cite{adelaide}. The physical $\kappa_v=3.706$ [n.m] is displayed by the full circle.}
    \label{fig:AB}
  \end{center}
\end{figure}

We first discuss the numerical results for schemes A and B. With most of the parameters of Eqs.(\ref{sa},\ref{sb}) determined from known physical quantities (resulting values are shown in Table \ref{table1}), it is clear that we have one unknown to fit in each case, $\kappa_{v,A}^0$ and $\kappa_{v,B}^0$. One quickly realizes that neither one of the two extrapolation functions provides an $m_\pi$-dependence that is compatible with the lattice data discussed in \cite{adelaide}. We therefore decide to fit $\kappa_{v,A}^0$ and $\kappa_{v,B}^0$ in such a way that we reproduce the physical $\kappa_v=3.706$ [n.m] \cite{PDG} for $m_\pi\rightarrow 0.138$ GeV. The result of scheme A (NLO HBChPT, Eq.(\ref{sa})) is shown in Fig.\ref{fig:AB} by the dashed curve, whereas scheme B (NLO SSE, Eq.(\ref{sb})) is represented by the dot-dashed curve. While both curves show a rather similar chiral limit behavior, they must be considered inadequate for $m_\pi>400$ MeV, as they even change sign in this region\footnote{Even when loosening our input provided by Table \ref{table1}, by allowing the (chiral limit) parameters of Eqs.(\ref{sa},\ref{sb}) to be slightly different from the physical values shown in Table \ref{table1}, the qualitative picture in Fig.\ref{fig:AB} does not change.}.  
In fact, the NLO extrapolations shown in Fig.\ref{fig:AB} provide such a poor $m_\pi$-dependence compared to the lattice data, that a (hypothetical) N$^n$LO higher order calculation in scheme A or B which might be more compatible with the data shown in Fig.\ref{fig:AB} would constitute such a large correction to the NLO result presented here, that one would have to worry about the ``convergence properties'' of the respective expansion scheme. To summarize our discussion on the numerics for schemes A and B we conclude that even future higher order calculations in these two approaches presumably can only describe the $m_\pi$-dependence of the isovector anomalous magnetic moment for $m_\pi<400$ MeV, making the comparison with lattice data impossible for the moment. We also note that the relativistic version of scheme A discussed in appendix \ref{B} seems to work well out to $m_\pi\approx 500$ MeV, as shown by the solid line in Fig.\ref{fig:AB}.  

\subsection{Numerical Analysis of Scheme C}

The situation is more complicated for the NLO result of scheme C (Eq.(\ref{sc})), where---at a given scale $\lambda$---we have, in principle, 7 
parameters to deal with: 
\begin{eqnarray}
\kappa_{v,C}^0,\,g_A^0,\,\fpi^0,\,c_A^0,\Delta_0,\,E_1^r(\lambda),\,c_V^0\;. \label{Cparameters}
\end{eqnarray} 
Ideally, we would fit these parameters to lattice QCD simulations of the isovector anomalous magnetic moment for $m_\pi\leq600$ MeV\footnote{We take this ``naive guess'' on the range of applicability of a leading-one-loop calculation based on the success of chiral effective field theory for predicting virtual Compton scattering cross sections on a nucleon at three momentum transfer $|\vec{q}|\sim 600$ MeV \cite{VCS}.}. However, at the moment the data situation does not allow fits in this parameter range \cite{adelaide}. 
Based on our estimate regarding  the applicability of leading-one-loop calculations in chiral effective field theories we could just stop here with our analysis and wait for future simulations at smaller quark masses before we continue to discuss the extrapolation curve given by Eq.(\ref{sc}). However, {\em we observe that the lattice data discussed in Ref.\cite{adelaide} are basically flat or at most weakly dependent on $m_\pi$} in the range 600 MeV $<\,m_\pi<$ 1 GeV$\approx\Lambda_\chi$ (c.f. Fig.\ref{fig:AB}). This observation leads us to the following hypothesis: 

{\em If the extrapolation function of a chiral effective field theory calculation contains sufficient quark mass dependent structures to yield such a ``plateau'' as suggested by the lattice data in this mass range, then the higher order corrections---though formally large $(\sim (m_\pi/\Lambda_\chi)^n)$ for $m_\pi>600$ MeV---are presumably small, as they cannot deviate much from the plateau to which the extrapolation curve is fitted.} 

The ``weak'' mass dependence of the lattice data for large quark masses in this view acts as ``boundary condition'' for the chiral extrapolation function, constraining the extrapolation to small quark masses more efficiently than expected from a chiral powercounting for individual polynomial structures.     
Obviously a necessary condition for this hypothesis to make sense is the requirement that the resulting extrapolation curve compares reasonably well with the empirical value for $\kappa_v$ at $m_\pi=138$ MeV. 

\begin{figure}[!htb]
  \begin{center}
    \includegraphics*[width=0.9\textwidth]{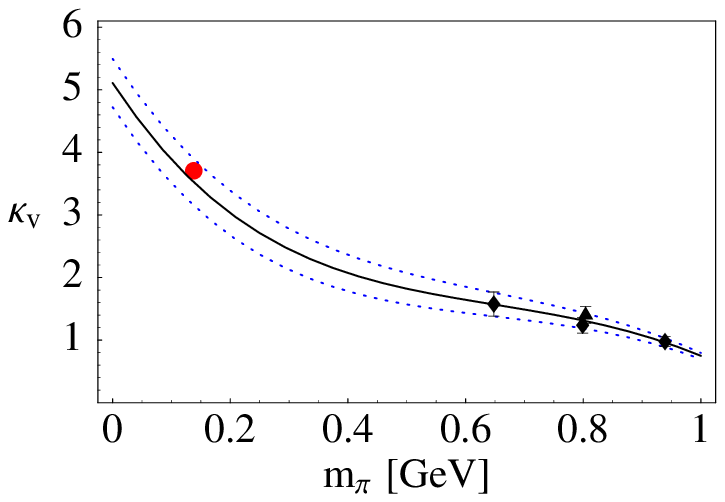}
    \caption{Pion mass dependence of the isovector anomalous magnetic moment in nuclear magnetons. The full curve denotes the NLO calculation in the modified expansion scheme C of Eq.(\ref{sc}). The lattice data are taken from Ref.\cite{adelaide}. The physical $\kappa_v=3.706$ [n.m] is displayed by the full circle.}
    \label{fig:C}
  \end{center}
\end{figure}

With this hypothesis in mind we now leave the secure realm of chiral effective field theory and attempt to find a set of numerical values for the parameters in list \ref{Cparameters} that is consistent with the lattice data of Ref.\cite{adelaide} in the range 600 MeV $<\,m_\pi<$ 1 GeV and still gives a meaningful ``prediction'' for $\kappa_v$ at $m_\pi=138$ MeV.
To pursue this program we employ the following philosophy: We determine the number of degrees of freedom that can be fixed from lattice data for effective pion masses $m_\pi<\Lambda_\chi\approx 1$ GeV. For the correlated $(\kappa,\,m_\pi)$ lattice data given in \cite{adelaide} it turns out that this number is three\footnote{There are actually four lattice points below 1 GeV discussed in \cite{adelaide}, based on the simulations of Refs.\cite{LDW,WDL}. However, the two points around $m_\pi=800$ MeV are only separated by 5 MeV, effectively only providing one degree of freedom for the fit.}. We then pick the corresponding number of couplings from list \ref{Cparameters} 
according to the following principles:
\begin{itemize}
\item[1)] Couplings for which we have no other physical information available. 
\item[2)] Couplings for which we expect a significant difference between the physical and the chiral limit value.
\end{itemize}
Based on the first argument we pick $\kappa_{v,C}^0$ and $E_1^r(\lambda)$ from list \ref{Cparameters}. This leaves us with one more parameter that we can constrain from lattice data. Based on argument two, we choose $c_V$, the leading coupling of the magnetic M1 $\gamma N\Delta$ transition in the chiral limit introduced in Eq.(\ref{eq:cv}). Indeed at the physical point one expects sizable complex-valued corrections to this transition from the pion cloud of the nucleon \cite{bss,ndelta} (as well as interference from three additional higher order couplings \cite{ndelta}). The remaining four couplings---$g_A^0,\,c_A^0,\,\fpi^0,\Delta_0$---are taken at their physical values $g_A,\,c_A,\,\fpi,\Delta$, as given by Table \ref{table1}, in complete analogy to the procedure in schemes A and B discussed in the previous section. The induced quark mass difference is again considered to be part of the N$^n$LO, $n\geq 2$ corrections in Eq.(\ref{sc}). Next we utilize the three lattice points of Ref.\cite{adelaide} together with the parameters given in Table \ref{table1} as input and generate---at a chosen scale\footnote{The chiral extrapolation curves shown in Figs.\ref{fig:AB}-\ref{fig:G} do of course not depend on the choice of $\lambda$.} $\lambda$---numerical estimates for the three unknown parameters $\kappa_{v,C}^0,\,E_1^r(\lambda)$ and $c_V$ shown in Table \ref{table2}. The resulting couplings are of reasonable size\footnote{It turns out that the radiative decay width of $\Delta$(1232) estimated from the fitted value of Table \ref{table2} is by a factor of 4 smaller than the number given in \cite{PDG}. Either this means that the properties of the magnetic $\gamma N\Delta$ transition are really substantially different in the chiral limit than at the physical point, or that there are significant higher order corrections which get lumped into an averaged number for $c_V$. This issue can only be decided in an N$^2$LO analysis \cite{massimiliano}.} and produce the chiral extrapolation curve of scheme C shown in Fig.\ref{fig:C}. The full curve is obtained by fixing the three couplings from the central values of the lattice data \cite{adelaide}, whereas the dotted curves indicate the error band resulting from the errors of the lattice data \cite{tony}. For large pion masses the curve reproduces---by construction---the nearly flat behavior suggested by the lattice data, whereas for low masses ($m_\pi<400$ MeV) one observes considerable curvature. Surprisingly the curves extrapolate rather well into the low mass region, as indicated by the full circle representing the physical $\kappa_v$. 

One can now ask the question ``how large is the isovector anomalous magnetic moment $\kappa_v$ in a world where the lattice pion mass is 138 MeV ?'' Given that this piece of information was not used in determining the parameters in Tables \ref{table1} and \ref{table2}, we obtain---via 
Eq.(\ref{sc})---the ``prediction''
\begin{eqnarray}
\kappa_v|_{m_\pi\rightarrow 138{\rm MeV}}=3.5\pm 0.4 \;\;{\rm [n.m.]}\;.\label{success}
\end{eqnarray}   
This does not seem to be a great achievement, given the experimental accuracy to which the anomalous magnetic moments of proton and neutron are known ($\kappa_v^{exp.}=3.706\dots$ [n.m.] 
\cite{PDG}). However, it is by no means obvious that our chiral extrapolation should come anywhere close to the experimental number in view of the rather large extrapolation range, the sizable error bars of the lattice data, the non-negligeable curvature required for a successful extrapolation to small quark masses and the large associated mass scales of the lattice pion. We therefore consider the result of Eq.(\ref{success}) a rather surprising success. Along the same lines one can also determine the isovector anomalous magnetic moment of the nucleon in the chiral limit. We obtain the prediction
\begin{eqnarray}
\kappa_v|_{m_\pi\rightarrow 0}=5.1\pm 0.4 \;\;{\rm [n.m.]}\;,\label{chilimit}
\end{eqnarray}   
{\it i.e.} we find a significant enhancement over the value at the physical point. We note that it is the coupling of the nucleon's pion cloud to the external electromagnetic field applied to probe the strength of the nucleon's magnetic moment, that is responsible for this reduction in $\kappa_v$ when one slowly increases the quark masses from 0 to about 8 MeV. Diagrammatically this effect is best displayed by diagram d) in Fig.\ref{fig:diags}, which provides the bulk of the effect and corresponds to the Caldi-Pagels term displayed in Eq.(\ref{sa}). 

\begin{table}
\begin{tabular}{|c|l|l|}\hline
Parameter & $\lambda=0.77$ GeV & $\lambda=1$ GeV \\\hline
$\kappa_{v,C}^0$ & $+ 5.1\mp 0.4$ & $+ 5.1\mp 0.4$\\
$c_V^0$ & $-\left(2.26\pm 0.06\right)$ GeV$^{-1}$ &  $-\left(2.26\pm 0.06\right)$ GeV$^{-1}$ \\
$E_1^r(\lambda)$ & $-\left(4.4\pm 0.1\right)$ GeV$^{-3}$ & $-\left(3.85\pm 0.1\right)$ GeV$^{-3}$ \\\hline
$\kappa_{s,C}^0$ & $- 0.11$ & $- 0.11$ \\
$E_2^r(\lambda)$ & $+ 0.074$ GeV$^{-3}$ & $+ 0.074$ GeV$^{-3}$ \\\hline
\end{tabular}
\caption{Values of the three isovector and two isoscalar parameters obtained from fitting to the lattice data of Ref.\cite{adelaide} for different values of the regularization scale $\lambda$, using as additional input the parameters displayed in Table \ref{table1}.\label{table2}}
\end{table}

Ultimately the hypothesis formulated above about the applicability of Eq.(\ref{sc}) into the realm of $m_\pi>600$ MeV due to suspected strong cancellations among the higher order corrections can only be tested once the N$^2$LO corrections are fully calculated and analyzed \cite{massimiliano}, but the non-trivial physical and chiral limit predictions given by Eqs.(\ref{success},\ref{chilimit}) look rather promising in this respect. We now proceed to the (numerical) chiral limit discussion, which will provide some insight into the dynamical origin of the successful extrapolation function of scheme C generated in this section. 

\subsection{Numerical Analysis of the Chiral Limit} \label{numchi}

In this section we discuss the interplay between the various contributions to $\kappa_v$ by looking at the chiral limit expansion of all three schemes discussed in section \ref{chiral}. With Tables \ref{table1} and \ref{table2} we obtain
\begin{eqnarray}
\kappa_v^A[n.m.]|_{NLO}&=&5.645-\frac{14.05}{GeV}\,m_\pi\nonumber\\
\kappa_v^B[n.m.]|_{NLO}&\approx&5.859-\frac{14.05}{GeV}\,m_\pi
                         +\frac{m_\pi^2}{GeV^2}\left(0.6480+5.780\log
                         \frac{m_\pi}{GeV}\right)\nonumber\\
                & &+\frac{m_\pi^4}{GeV^4}\left(16.95+19.66\log
                   \frac{m_\pi}{GeV}\right)+\ldots\nonumber\\
\kappa_v^C[n.m.]|_{NLO}&\approx&5.109-\frac{14.05}{GeV}\,m_\pi
                       +\frac{m_\pi^2}{GeV^2}\left(36.63+5.780\log
                         \frac{m_\pi}{GeV}\right)\nonumber\\
                & &-61.63\frac{m_\pi^3}{GeV^3}
                   +\frac{m_\pi^4}{GeV^4}\left(24.43-34.61\log
                   \frac{m_\pi}{GeV}\right)+\dots \;,\label{eq:chiralnumerik}
\end{eqnarray}
which leads to the following observations:
\begin{itemize}
\item[1)] These approximate formulae only hold for pion masses below 400 MeV. For detailed numerical studies one should use the formulae of Eq.(\ref{sc}) which contain the full analytic structure.
\item[2)] These formulae display Taylor coefficients of a series in $m_\pi^n$. There is no hierarchy in these numbers in the sense that coefficients for small values of n are larger than the ones for high values of n. As already explained in
section \ref{chiral}, these coefficients just arise from Taylor expanding the logarithms of Eqs.(\ref{sa},\ref{sb},\ref{sc}). All the structures---except for the quark mass independent leading terms--- displayed in Eq.(\ref{eq:chiralnumerik}) are in fact {\em part of the same chiral order} (NLO) in their respective expansion schemes A, B and C. It is therefore not meaningful to look for ``convergence'' in this representation of the quark mass expansion of the magnetic moments. Convergence will be studied by calculating the N$^2$LO corrections to Eqs.(\ref{sa},\ref{sb},\ref{sc}), Taylor expanding the full result as done here and then {\em comparing by what amount the individual Taylor coefficients displayed in Eq.(\ref{eq:chiralnumerik}) have changed when one moves from NLO to N$^2$LO accuracy}. At present the N$^2$LO analyses in schemes B and C do not exist yet \cite{massimiliano}. As indicated in section \ref{calculation} they involve several new unknown couplings, therefore a detailed evaluation can only take place if more lattice data (preferably at smaller quark masses) become available in the correlated $(\kappa,\,m_\pi)$ representation employed here.
\item[3)] Eq.(\ref{eq:chiralnumerik}) gives a good idea why scheme C is so much more effective than scheme B with its ``naive power counting''. The extra structure $\sim m_\pi^3$ actually carries a rather large coefficient\footnote{For a discussion of these structures in leading-one-loop relativistic baryon ChPT we refer to Appendix \ref{B}.} and also the terms $\sim m_\pi^2$ are significantly enhanced in C. As can be seen from Eq.(\ref{eq:chiral}), both structures are intimately connected with the isovector $N\Delta$ transition governed by $c_V$ and the induced counterterm $E_1$. Given that these two couplings have such a strong impact on the quadratic and the cubic term in the chiral expansion, our proposal of Eq.(\ref{eq:cv}) seems well justified.
\end{itemize}
With these remarks we close our discussion on the chiral extrapolation function of the isovector anomalous magnetic moment and move on to the isoscalar sector.

\section{Isoscalar Anomalous Magnetic Moment}\label{isoscalar}
\setcounter{equation}{0}


In contrast to the isovector anomalous magnetic moment, there is hardly any quark mass dependence for the isoscalar anomalous
magnetic moment of the nucleon at NLO in chiral effective field theory calculations. In the three schemes discussed above one finds
\begin{eqnarray}
\kappa_s^A&=&\kappa_{s,\,A}^0+0+N^2LO \nonumber\\
\kappa_s^B&=&\kappa_{s,\,B}^0+0+N^2LO \nonumber\\
\kappa_s^C&=&\kappa_{s,\,C}^0-8\,E_2\,M\,m_\pi^2+N^2LO\;.\label{eq:kappascalar}
\end{eqnarray}
Only scheme $C$ shows any dependence on the quark masses to this order \cite{Nstar}. We observe that this quark mass dependence is not related to chiral dynamics but solely arises from an internal quark mass dependence of the core/bare spin 1/2 nucleon. Its origin (and the strength of the associated counterterm $E_2$) is therefore outside the range of the effective field theory.
 
\begin{figure}[!htb]
  \begin{center}
    \includegraphics*[width=0.9\textwidth]{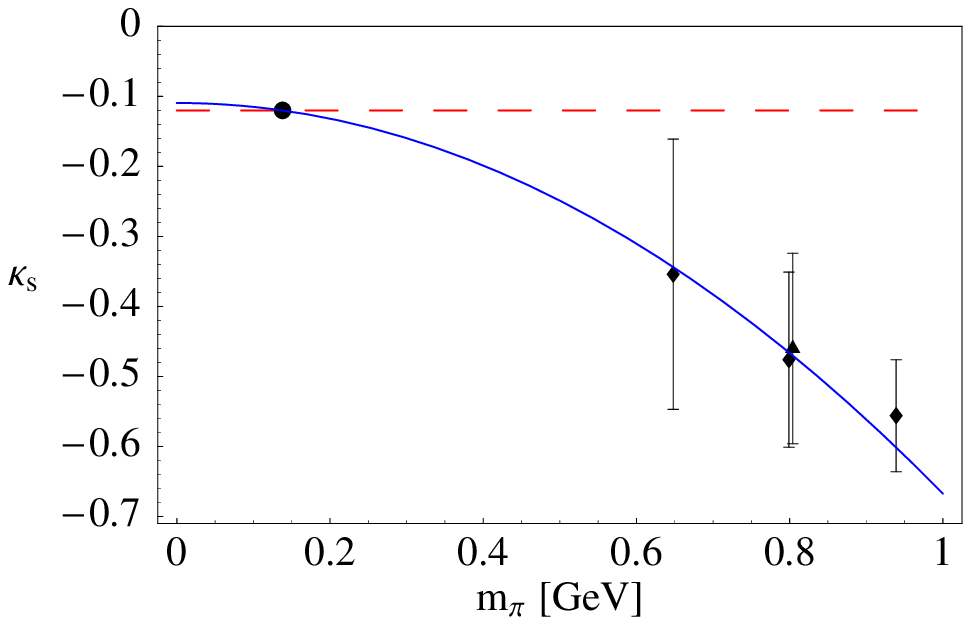}
    \caption{Pion mass dependence of the isoscalar anomalous magnetic moment in nuclear magnetons. The full curve represents the suggested pion mass dependence of the modified expansion scheme C of Eq.(\ref{eq:kappascalar}). The lattice data are taken from Ref.\cite{adelaide}. The physical $\kappa_s=-0.1202$ [n.m] is displayed by the full circle.}
    \label{fig:D}
  \end{center}
\end{figure}

An extrapolation of the isoscalar moment data to smaller quark masses analogous
to the one in the isovector sector is therefore unreliable at this order. 
Moreover, due to the smallness of the isoscalar anomalous magnetic moment and the correspondingly large
error bars of the lattice simulation reported in \cite{adelaide}, we must conclude that at present extrapolations in the isoscalar sector
are not feasible. Fig.\ref{fig:D} summarizes the lattice data situation for pion masses below 1 GeV. The curve shown in this figure is obtained by
fitting $\kappa_{s\,C}^0$ and $E_2$ to the physical value and to the lattice point at $m_\pi=800$ MeV. The resulting fit parameters are given in Table \ref{table2}. Our result in the isoscalar sector therefore is not meant as an extrapolation but at the moment
merely serves as a ``best guess'' (within scheme $C$) for the quark mass dependence in $\kappa_s$.  Nevertheless we have to utilize this result 
later
when we compare with the Pade fits of the Adelaide group for the magnetic moments of proton and neutron.

Finally, we briefly comment on the leading quark mass dependence to the isoscalar anomalous magnetic moment of the nucleon, that {\em does} arise from chiral dynamics. It is known that at the one-loop level one obtains a contribution if one extends the theory to the case of three active flavors---SU(3) Baryon ChPT---with quark masses $(\hat{m},m_s)$. For the leading non-analytic quark mass dependence in a theory with only octet mesons and octet baryons as active degrees of freedom---corresponding to scheme A in the SU(2) sector---one finds \cite{Pagels}
\begin{eqnarray}
\kappa_s^{SU(3)}=\kappa_s^0-\frac{M_N\,m_K}{24\pi F_\pi^2}\left(5D^2-6 D F+9F^2\right)+N^2LO\;,\label{eq:kaons}
\end{eqnarray}
with
\begin{eqnarray}
m_K^2=\left(\hat{m}+m_s\right)B_0\left\{1+{\cal O}({\cal M})\right\}
\end{eqnarray}
denoting the kaon mass (squared). We note that this contribution arises from diagram d) in Fig.\ref{fig:diags} if one allows the complete baryon and meson octet as possible intermediate states. $F_\pi=(f_\pi+f_K)/2$ denotes the average of the pion and kaon decay constants to this order and $F,\,D$ are the SU(3) axial coupling constants \cite{JM} with the constraint $g_A=F+D$. We also note that for the case of three active flavors the isoscalar anomalous magnetic moment of the nucleon arises from two short distance counterterms ({\it e.g.} see the Lagrangian discussed in \cite{Jenkins}). Given that the present lattice data in the isoscalar sector as shown in Fig.\ref{fig:D} are not sufficiently accurate, we do not embark on a numerical study of Eq.(\ref{eq:kaons}) or its combination with Eq.(\ref{eq:kappascalar}).


\section{Comparison to Pade Approximants and the Quark Model}\label{Pade}
\setcounter{equation}{0}

\begin{figure}[!htb]
  \begin{center}
    \includegraphics*[width=0.9\textwidth]{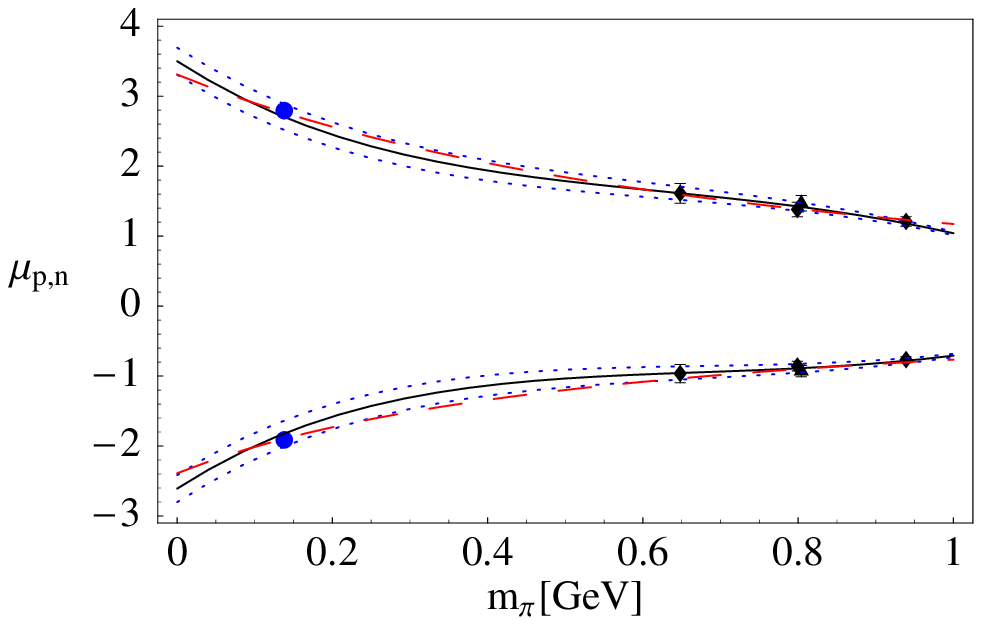}
    \caption{Pion mass dependence of the magnetic moments of proton (upper curves) and neutron (lower curves) in nuclear magnetons. The full curve represents the best fit in the modified expansion scheme, whereas the dashed curve denotes the Pade extrapolation formula Eq.(\ref{eq:pade}). Our error estimate is given by the dotted curves. The lattice data are taken from Ref.\cite{adelaide}. The physical values $\mu_p=2.793$ [n.m], $\mu_n=-1.913$ [n.m] are displayed by the full circles.}
    \label{fig:E}
  \end{center}
\end{figure}

In 1998 the Adelaide group suggested \cite{adelaide} a simple parameterization for the quark mass dependence of the nucleons' magnetic moments
based on Pade approximants. 
For their ``best fit''\footnote{The Adelaide group also includes lattice points above 1 GeV pion mass in their
analysis and obtained a good fit throughout the whole region in $m_\pi$.} they used the following functional dependence,
\begin{eqnarray}
\mu_{p,n}&=&\frac{\mu^0_{p,n}}{1+\alpha_{p,n}\, m_\pi+\beta_{p,n}\, m_\pi^2}\;, \label{eq:pade}
\end{eqnarray}
with values of the parameters given in Table \ref{table3}. 
\begin{table}
\begin{tabular}{|c|l||c|l|}\hline
Proton & Value & Neutron & Value \\\hline
$\mu_p^0$ & 3.31 [n.m.] & $\mu_n^0$ & -2.39 [n.m.] \\
$\alpha_p$ & 1.37 GeV$^{-1}$ & $\alpha_n$ & 1.85 GeV$^{-1}$ \\
$\beta_p$ & 0.452 GeV$^{-2}$ & $\beta_n$ & 0.271 GeV$^{-2}$ \\\hline
\end{tabular}
\caption{Parameters used in the Pade fit of the magnetic moments, taken from Ref.\cite{adelaide}.\label{table3}}
\end{table}
In Fig.\ref{fig:E} we show this Pade fit as the dashed curve, whereas our combined central value result in scheme C
\begin{eqnarray}
\mu_p^C&=&1+(\kappa_s^C+\kappa_v^C)/2\;,\nonumber\\
\mu_n^C&=&(\kappa_s^C-\kappa_v^C)/2\;,\label{eq:magmoment}
\end{eqnarray}
is represented by the full curve, utilizing the parameters given in Tables \ref{table1} and \ref{table2} for $\kappa_v^C$ from Eq.(\ref{sc}) and $\kappa_s^C$ from Eq.(\ref{eq:kappascalar}). Surprisingly,
both parameterizations agree quite well and are certainly compatible within the present error band. Minor differences can be
found in the curve for the neutron and for really small pion masses near the chiral limit, but at present both parameterizations are indistinguishable due to the sizable extrapolation errors originating from the lattice error bars \cite{tony}. Both parameterizations of the chiral corrections to
the magnetic moments indicate that for small quark masses
approaching the ``physical'' values, there is indeed a substantial curvature with respect to $m_\pi$ beyond the leading-non-analytic Caldi-Pagels term of Eq.(\ref{sa}). We note again that, a priori, there is no reason to expect that a simple Pade ansatz as in Eq.(\ref{eq:pade}) captures all the important chiral physics, especially in view of potentially large logarithmic terms generated by the pion-cloud dynamics. Judging from the chiral limit result of our chiral extrapolation C displayed in Eq.(\ref{eq:chiralnumerik}) the numerical coefficients in front of the $m_\pi^{2n}\log m_\pi$ structures are either small (c.f. n=1) or get canceled by corresponding $m_\pi^{2n}$ polynomial structures (c.f. n=2). The dynamical origin of this logarithmic suppression is not known at the moment. Nevertheless, we observe
that our microscopic calculation agrees well with the extrapolation formula Eq.(\ref{eq:pade}) for the proton and neutron magnetic moments.

\begin{figure}[!htb]
  \begin{center}
    \includegraphics*[width=0.9\textwidth]{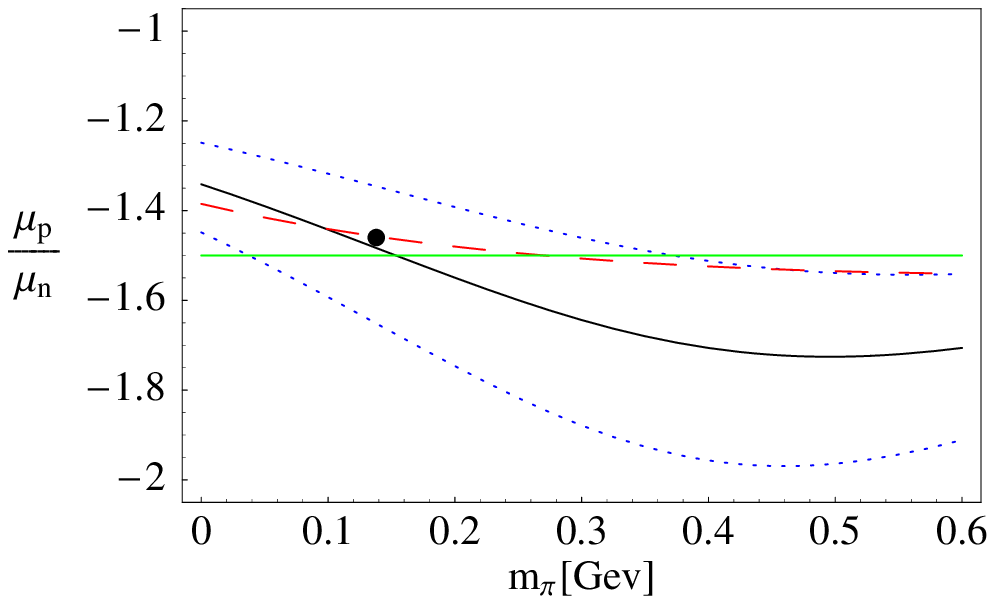}
    \caption{Pion mass dependence of the ratio of the magnetic moments. The physical value $\mu_p/\mu_n=-1.46$ is denoted by the full circle, whereas the SU(6) quark model prediction of -3/2 is given by the horizontal straight line. The chiral extrapolation result based on Eq.(\ref{eq:magmoment}) is given by the solid curve, with extrapolation errors indicated by the dotted lines.  The corresponding curve based on the Pade formula of Eq.(\ref{eq:pade}) is shown as the dashed line.}
    \label{fig:F}
  \end{center}
\end{figure}

It is instructive to compare the quark mass dependence of the ratio of magnetic moments with SU(6) quark model predictions. In Ref.\cite{LTY} it was noted that for a pion mass of $\sim 240$ MeV the proton to neutron ratio $\mu_p/\mu_n$ would yield the quark model prediction of -3/2, leading to the conclusion that the good agreement between the quark model---which knows nothing of the light current quark masses discussed here---and the experimental ratio $\mu_p/\mu_n=-1.46$ is accidental. In Fig.\ref{fig:F} we show that our parameterization for the ratio of the magnetic moments given via Eq.(\ref{eq:magmoment}) (full curve) follows the trend of the Pade formula (dashed curved) \cite{LTY}. Our best fit curve already gives a ratio of -3/2 for $m_\pi\sim 150$ MeV, albeit the error band of our extrapolation (dotted curves) arising from the lattice data is sizable in this quark mass regime, pointing again to the need for lattice data at smaller quark masses to set stronger constraints on the chiral extrapolation. The restoration of the quark model results in the limit of very heavy quark masses discussed in \cite{LTY} is beyond the realm of applicability of chiral effective field theory, so we will not discuss it here.

\begin{figure}[!htb]
  \begin{center}
    \includegraphics*[width=0.9\textwidth]{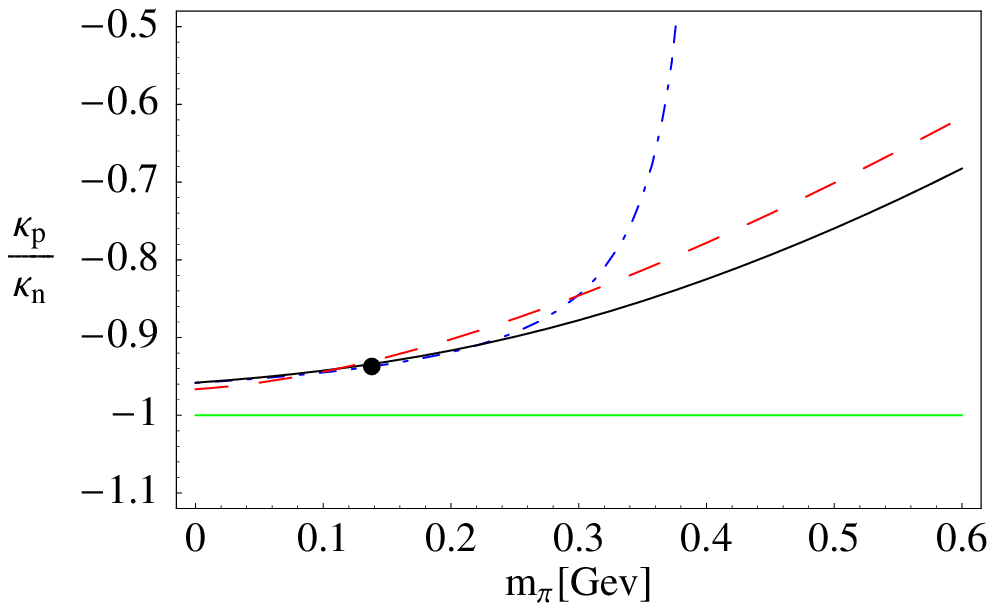}
    \caption{Pion mass dependence of the ratio of the anomalous magnetic moments of proton and neutron. The SU(6) quark model prediction of -1 is given by the straight horizontal line. The chiral extrapolation curve based on scheme C is given by the solid and the Pade fit by the dashed curve. The physical value $\kappa_p/\kappa_n=-0.94$ is denoted by the full circle. The dot-dashed curve shows the dependence of this ratio on the leading-non-analytic quark mass term only, corresponding to scheme A discussed in the text.}
    \label{fig:G}
  \end{center}
\end{figure}

While the $\mu_p/\mu_n$ ratio shown in Fig.\ref{fig:F} allows extrapolation curves to lie near the SU(6) quark model prediction at least for certain values of light quark masses, this is not the case for the ratio of the anomalous magnetic moments shown in Fig.\ref{fig:G}. The SU(6) quark model value $\kappa_p/\kappa_n=-1$ is not reached by any of the two extrapolation curves in the small quark mass regime. However, it is interesting to note that this ratio seems to be rather insensitive to quark mass effects beyond the leading-non-analytic (Caldi-Pagels) term given in Eq.(\ref{sa}) for pion masses up to 250 MeV, indicated by the dot-dashed curve in Fig.\ref{fig:G} corresponding to scheme A of Eqs.(\ref{sa},\ref{eq:kappascalar}). Whereas in the other observables (e.g. see Figs.\ref{fig:C},\ref{fig:E}) the leading-non-analytic term is not even sufficient to extrapolate from the chiral limit to the physical pion mass, the ratio $\kappa_p/\kappa_n$ seems well suited for future chiral extrapolations of lattice data with small quark masses. With this observation we move on to a brief discussion regarding the influence of quenching.
.


\section{Effects of Quenching}
\setcounter{equation}{0}


The lattice data on magnetic moments shown in Figs.\ref{fig:AB}-\ref{fig:G} 
have been obtained in
the quenched approximation of QCD, i.e. with the (loop-) effects of sea quarks
effectively suppressed. A fully consistent treatment of the chiral
extrapolation should therefore be adapted to this situation, in the sense that
the effects of quenching should also be taken into account appropriately within
a chiral effective field theory. Such a framework has been developed in the
past decade, called Quenched Chiral Perturbation Theory (QChPT)
\cite{Bernard}. When performing chiral extrapolations of quenched lattice data,
one does not only have to take into account that the prefactors of pion-mass
dependent terms could be different in QChPT---even the chiral 
singularity structure can be different between
``Quenched'' and ``Full'' QCD. Specifically, for the case of the nucleon
magnetic moments \cite{qmoment} of interest here, this means that there are
terms proportional to $const.\times\log m_\pi$ which do not exist in ``Full''
QCD. The chiral expansion of $\kappa_v$ of Eq.(\ref{eq:chiral}) demonstrates
that the leading logarithmic pion-mass dependence is proportional to
$m_\pi^2\times\log m_\pi$ in ``Full'' QCD.
The presence of a new chiral singularity like
$const.\times\log m_\pi$ in ``Quenched'' QCD indicates that its low energy properties---not to speak of its chiral limit---can differ qualitatively from what we know in hadron phenomenology.

While ``Quenched'' QCD is interesting in its own right, with the comments just
made one might arrive at the pessimistic conclusion that not much can be
learned about hadron properties in ``Full'' QCD via extrapolations of quenched
lattice data. However, the loop-effects of the sea-quarks---which are missing in ``Quenched'' QCD--- get strongly
suppressed for large quark masses, leading one to expect that ``Quenched'' and
``Full'' QCD do not differ much in the ``heavy quark regime''. In addition,
large quark masses mean that one is ``far away'' from chiral
singularities which dominate the chiral limit behavior of the quantities of
interest. For the time being we can therefore identify quenched lattice data
approximately with ``Full'' QCD {\em for effective pion masses above 600 MeV}. This assumption is further supported by three observations: 
\begin{itemize}
\item[1)] Available quenched lattice data on the magnetic moments of the nucleon for pion masses in the range 0.6 GeV $< m_\pi < $ 1.5 GeV show a rather moderate curvature ({\it e.g.} see Fig.5 in Ref.\cite{adelaide}), suggesting a negligeable effect of the extra chiral singularity in this mass region. 
\item[2)] The difference between the Pade extrapolation curve Eq.(\ref{eq:pade}) discussed in section \ref{Pade} and a similar formula that explicitly includes the leading effects of quenching has been reported to be small \cite{private}. For pion masses above 600 MeV the two analyses lie within the error bars of the lattice data.
\item[3)] No significant differences between quenched and fully dynamical simulations for $m_\pi>600$ MeV have been reported for a variety of nucleon structure properties, see for example the recent study of moments of nucleon quark distributions by the LHPC/SESAM collaboration \cite{SESAM}. 
\end{itemize}
In essence, while ``Quenched'' QCD does have a chiral limit different from ``Full'' QCD we can accept the given lattice data points for proton and neutron magnetic moments at $m_\pi>600$ MeV as if they were ``unquenched'', within their uncertainties. An extrapolation using standard (rather than quenched) effective chiral field theory seems therefore justified to provide the correct chiral extrapolation to small quark masses as well as the proper chiral limit, at least for our present purpose.


\section{Conclusion and Outlook}
\setcounter{equation}{0}


The present analysis has pointed out the feasibility of systematic chiral
extrapolations of nucleon magnetic moments from lattice QCD, down to the range
of realistic light quark masses where comparisons with the actual observables
can be made. An important element in this discussion is the treatment of the
$\Delta$(1232) isobar as an {\em explicit} degree of freedom in view of its
important role in the magnetic structure of the nucleon. In our approach it is
this feature which produces the important non-analytic quark mass dependencies of the magnetic moments {\em beyond} the well-known Caldi-Pagels term proportional to $m_\pi$.

Our resulting extrapolation is remarkably close to the Pade approximant
parameterization of the Adelaide group. A sign of caution should be added,
however. The existing (quenched) data terminate around an effective pion mass
of 0.65 GeV, corresponding to $u$- and $d$-quark masses $\hat{m}_{u,d}\sim 200$
MeV and larger. Expanding one-loop chiral effective field theory to such large
quark mass scales has its inherent uncertainties which induce a substantial
error in the extrapolation down to small quark masses. In fact, it is rather
surprising that our NLO extrapolation curve ends up near the physical value for
$\kappa_v$, given that we only use the input from these rather large mass
scales. Certainly the N$^2$LO corrections in schemes $A,\,B$ and $C$ (as well as in the relativistic approach discussed in Appendix \ref{B}) 
need to be analyzed systematically to judge the stability of
the extrapolation \cite{massimiliano}. Future (partially) unquenched lattice
simulations aiming for effective pion masses around 300 MeV are also expected
to reduce the extrapolation uncertainties significantly. According to our
results deviations from the nearly linear trend seen so far in the data below 1
GeV pion mass should then become visible in the lattice data. However,
extrapolating the results of such intermediate mass scale simulations down to
the ``physical'' pion mass of 138 MeV will presumably require precision
calculations within (partially) quenched chiral effective field theories, in
order to get control over the magnitude of the effects of (partial) quenching in the simulations. Pioneering studies in this direction have already been performed and look promising \cite{pq}.

As a final remark we note that future lattice simulations (and the associated chiral extrapolations) of magnetic moments and related nucleon properties should preferentially be done in the isovector/isoscalar basis, as the two channels show quite different patterns of quark mass dependence. 


\bigskip

\bigskip

\bigskip

\bigskip

\begin{center}
{\bf Acknowledgments}
\end{center}

TRH would like to acknowledge the hospitality of ECT* in Trento where a large part of this work was done as well as the hospitality of the UW Nuclear Theory group in Seattle where it was finally completed . The authors also acknowledge
helpful discussions with M. G{\" o}ckeler, N. Kaiser, D.B. Leinweber, M.J. Savage, A. Sch{\" a}fer, A.W. Thomas and M. Vanderhaeghen. 
This work was supported in part by BMBF and DFG.


\appendix

\section{Amplitudes}\label{A}
\setcounter{equation}{0}

Here we present the results for  the 11 leading-one-loop ({\it i.e.} ${\cal O}(\epsilon^3)$ diagrams shown in Fig.\ref{fig:diags} which can contribute to the anomalous magnetic moment of a nucleon of mass $M_N$. The Lagrangians needed for this calculation are discussed in section \ref{Lagrangians}. We work in the Breit-frame and choose the velocity vector $v^\mu=(1,0,0,0)$. With $S^\mu$ denoting the Pauli-Lubanski spin-vector and $\epsilon^\mu$ denoting the polarization 4-vector of an incoming photon with 4-momentum $q^\mu$ one finds
\begin{eqnarray}
Amp_{3a}&=&\frac{i e}{2 M_N}\,\bar{u}_v\left[S\cdot\epsilon,S\cdot q\right]\left\{\kappa_s^0+\kappa_v^0\tau^3-4M_N\Delta D_2-4M_N\Delta D_1\tau^3-8M_Nm_\pi^2E_2\right.\nonumber\\
& &\left.-8M_Nm_\pi^2E_1\tau^3-4M_N\Delta^2 D_4-4M_N\Delta^2 D_3\tau^3+{\cal O}(\epsilon^4)\right\}u_v \nonumber\\
Amp_{3b}&=&Amp_{3c}=0+{\cal O}(\epsilon^4) \nonumber\\
Amp_{3d}&=&\frac{i e}{2 M_N}\,\bar{u}_v\left[S\cdot\epsilon,S\cdot q\right]\left\{-\frac{g_A^2M_Nm_\pi}{4\pi\fpi^2}+{\cal O}(\epsilon^4)\right\}\tau^3u_v\nonumber\\
Amp_{3e}&=&0+{\cal O}(\epsilon^4)\nonumber\\
Amp_{3f}&=&Amp_{3g}=0+{\cal O}(\epsilon^4) \nonumber\\
Amp_{3h}&=&i\,\frac{c_A^2 e}{\fpi^2}\,\frac{8}{3(d-1)}\,\bar{u}_v\tau^3\left[S\cdot\epsilon,S\cdot q\right]u_v\,\frac{\partial}{\partial m_\pi^2}\,J_2\left(-\Delta,m_\pi^2\right)+{\cal O}(\epsilon^4)\nonumber\\
&=&\frac{i e}{2 M_N}\,\bar{u}_v\left[S\cdot\epsilon,S\cdot q\right]\left\{\frac{2c_A^2\Delta M_N}{9\pi^2\fpi^2}\left(16\pi^2L+\log\frac{2\Delta}{\lambda}\right)-\frac{5c_A^2\Delta M_N}{27\fpi^2\pi^2}\right.\nonumber\\
& &\left.+\frac{2c_A^2\Delta M_N}{9\pi^2\fpi^2}\left[\log\left(\frac{m_\pi}{2\Delta}\right)+\frac{\sqrt{\Delta^2-m_\pi^2}}{\Delta}\log R\right]+{\cal O}(\epsilon^4)\right\}\tau^3u_v\nonumber\\
Amp_{3i}&=&0+{\cal O}(\epsilon^4)\nonumber\\
Amp_{3j}&=&Amp_{3k}\nonumber\\
&=&i\,\frac{c_Ag_Ac_Ve}{\fpi^2\Delta}\frac{8(d-3)}{3(d-1)}\,\bar{u}_v\tau^3\left[S\cdot\epsilon,S\cdot q\right]u_v\left\{J_2(-\Delta,m_\pi^2)-J_2(0,m_\pi^2)\right\}+{\cal O}(\epsilon^4)\nonumber\\
&=&\frac{i e}{2 M_N}\,\bar{u}_v\left[S\cdot\epsilon,S\cdot q\right]\left\{
-\frac{4c_Ac_Vg_AM_N\Delta^2}{27\pi^2\fpi^2}\left[16\pi^2L+\log\frac{2\Delta}{\lambda}\right]\right.\nonumber\\
& &+\frac{2c_Ac_Vg_AM_Nm_\pi^2}{9\pi^2\fpi^2}\left[16\pi^2L+\log\frac{2\Delta}{\lambda}\right]
+\frac{2c_Ac_Vg_AM_N\Delta^2}{81\pi^2\fpi^2}
+\frac{2c_Ac_Vg_AM_Nm_\pi^3}{27\pi\Delta\fpi^2}\nonumber\\
& &-\left.\frac{4c_Ac_Vg_AM_N\Delta^2}{27\pi^2\fpi^2}\left[\left(1-\frac{m_\pi^2}{\Delta^2}\right)^{3/2}\log(R)+\left(1-\frac{3m_\pi^2}{2\Delta^2}\right)\log\left(\frac{m_\pi}{2\Delta}\right))\right]
\right\}\tau^3u_v \nonumber\\
& &+{\cal O}(\epsilon^4)\;.\label{amplitudes}
\end{eqnarray}
Explicit expressions for the function $J_2$ are given in \cite{Compton}. We evaluate the amplitudes in $d$-dimensions with induced regularization scale $\lambda$. Any ultraviolet divergences appearing in the limit $d\rightarrow 4$ are subsumed in 
\begin{eqnarray}
L=\frac{\lambda^{d-4}}{16\pi^2}\left[\frac{1}{d-4}+\frac{1}{2}\left(\gamma_E-1+\ln 4\pi\right)\right],
\end{eqnarray}
where $\gamma_E$ denotes the Euler-Mascharoni constant.

To simplify the calculation we have utilized the electromagnetic gauge-condition $v\cdot\epsilon=0$. Amplitudes e) and i) are zero due to this choice of gauge, whereas the null result of amplitudes b), c), f) and g) follows from the Pauli-Lubanski condition $S\cdot v=0$. To this order in the calculation the non-zero results therefore arise solely from amplitudes a), d), h), j) and k). However, diagrams b), c), e), f), g) i) will start contributing\footnote{Note that at ${\cal O}(\epsilon^4)$ one also has to take into account tadpole topologies and wavefunction renormalization graphs, which are not displayed in Fig.\ref{fig:diags}.} at ${\cal O}(\epsilon^4)$ with a strength depending on our choice of $v^\mu$.

\section{$\kappa_v$ to NLO in Relativistic Baryon ChPT}\label{B}
\setcounter{equation}{0}

Expansion schemes A, B and C discussed in the main text are non-relativistic approaches based on the ``Heavy Baryon'' method of \cite{JM}. Recently, the relativistic one loop analysis of the nucleon magnetic moments presented in \cite{Gasser} has been updated in Ref.\cite{KM}, employing a new regularization scheme \cite{becher} which overcomes the large renormalization effects $\sim M^n$ typically plaguing relativistic Baryon ChPT in standard dimensional regularization \cite{Gasser}. At leading-one-loop order (i.e. NLO) in a relativistic chiral effective field theory with pions and nucleons as the degrees of freedom one obtains \cite{KM}
\begin{eqnarray}
\kappa_v^{IR}&=&c_6-\frac{g_A^2M^2x^2}{16\pi^2\fpi^2}\left\{\frac{3x^2}{2}+\frac{2\left(8-13x^2+3x^4\right)}{x\sqrt{4-x^2}}\arccos\left[-\,\frac{x}{2}\right]+2\left(7-3x^2\right)\log x\right\}\nonumber\\
& &+N^2LO \;,\label{eq:relat}
\end{eqnarray}
with $x=m_\pi/M$. This result arises from diagrams a) -e) in Fig.\ref{fig:diags}, as well as tadpole and wavefunction renormalization. We are interested in this result because relativistic corrections to the Caldi-Pagels term given in Eq.(\ref{sa}) can also generate structures $\sim m_\pi^3$, which are claimed to be important for the fitting of the lattice data (c.f. section \ref{numchi}). To discuss this structure we perform the chiral limit expansion and obtain
\begin{eqnarray}
\kappa_v^{IR}|_{NLO}&\approx&c_6-\frac{g_A^2M}{4\pi\fpi^2}\,m_\pi
                     -m_\pi^2\left[\frac{g_A^2}{4\pi^2\fpi^2}
                     +\frac{7g_A^2}{8\pi^2\fpi^2}\log\frac{m_\pi}{M}
                     +N^2LO\right]\nonumber\\
                    &       &+m_\pi^3\left[\frac{3g_A^2}{8\pi M\fpi^2}
                     +N^2LO\right]+\dots\;,
\end{eqnarray}
which agrees with Ref.\cite{Gasser}, where terms up to $m_\pi^2\log m_\pi$ were discussed. Indeed one also finds a structure $\sim m_\pi^3$ in the relativistic approach, as expected. In order to judge its importance we
utilize the parameters of Table \ref{table1} and fix the unknown constant $c_6$ to reproduce $\kappa_v=3.706$ [n.m.] for $m_\pi\rightarrow 138$ MeV. We obtain the solid curve shown in Fig.\ref{fig:AB}. Its $m_\pi$-dependence is superior to scheme A or B when compared to the lattice data of Ref.\cite{adelaide}. However, the NLO relativistic result of Eq.(\ref{eq:relat}) breaks down for $m_\pi\approx 600$ MeV. Returning to the chiral limit discussion we find
\begin{eqnarray}
\kappa_v^{IR}[n.m.]|_{NLO}&\approx&5.068-\frac{14.05}{GeV}\,m_\pi
                       +\frac{m_\pi^2}{GeV^2}\left(-5.814-16.67\log
                         \frac{m_\pi}{GeV}\right)\nonumber\\
                & &+23.90\frac{m_\pi^3}{GeV^3}
                   +\frac{m_\pi^4}{GeV^4}\left(6.364+8.104\log
                   \frac{m_\pi}{GeV}\right)+\dots \;.
\end{eqnarray}
Comparing this result to Eq.(\ref{eq:chiralnumerik}) we conclude that the relativistic approach like scheme C contains all allowed quark mass structures, albeit with different (in general smaller) coefficients. We note that the structures $\sim m_\pi^2,\,m_\pi^3$ have a different sign in the relativistic approach than in Eq.(\ref{eq:chiralnumerik}). At N$^2$LO in relativistic Baryon ChPT some effects related to $\Delta$(1232) are implicitly incorporated\footnote{A consistent scheme to introduce $\Delta$(1232) as an explicit degree of freedom also in the relativistic approach is in preparation. V. Bernard, private communication.} in the higher order counterterms of the delta free theory. It will be interesting to see whether at that order the relativistic approach can be extended to pion masses above 600 MeV. A N$^2$LO stability analysis of schemes A, B, C and of the relativistic approach is in preparation 
\cite{massimiliano}.


\end{document}